\documentclass[12pt, pre,aps, showpacs,preprint]{revtex4}

\pdfoutput=1
\usepackage{amsmath}
\usepackage{amsfonts}
\usepackage[pdftex]{graphicx}
\usepackage{mathrsfs}
\usepackage{rotating}
\usepackage{setspace}
\usepackage{supertabular}
\usepackage{multirow}
\usepackage{amsthm}
\usepackage{subfigure}

\begin{document}
\title{Densest local sphere-packing diversity: General concepts and application to two dimensions}
\author{Adam B. Hopkins and Frank H. Stillinger}
\affiliation{Department of Chemistry, Princeton University, Princeton, New 
Jersey 08544}

\author{Salvatore Torquato}
\affiliation{Department of Chemistry, Princeton Institute for the Science and 
Technology of Materials, Program in Applied and Computational Mathematics, 
Princeton Center for Theoretical Science, Princeton University, Princeton, 
New Jersey 08544 \\ School of Natural Sciences, Institute for Advanced Study, 
Princeton, New Jersey 08544}

\begin{abstract}
The densest local packings of $N$ identical nonoverlapping spheres within a radius $R_{min}(N)$ of a fixed central sphere of the same size are obtained using a nonlinear programming method operating in conjunction with a stochastic search of configuration space. Knowledge of $R_{min}(N)$ in $d$-dimensional Euclidean space $\mathbb{R}^d$ allows for the construction both of a realizability condition for pair correlation functions of sphere packings and an upper bound on the maximal density of infinite sphere packings in ${\mathbb R}^d$. In this paper, we focus on the two-dimensional circular disk problem. We find and present the putative densest packings and corresponding $R_{min}(N)$ for selected values of $N$ up to $N=348$ and use this knowledge to construct such a realizability condition and upper bound. We additionally analyze the properties and characteristics of the maximally dense packings, finding significant variability in their symmetries and contact networks, and that the vast majority differ substantially from the triangular lattice even for large $N$. Our work has implications for packaging problems, nucleation theory, and surface physics.
\end{abstract}

\maketitle

\section{Introduction}

A packing is defined as a set of nonoverlapping objects arranged in a space of given dimension. Packings of identical nonoverlapping spheres in $d$-dimensional Euclidean space ${\mathbb R}^d$ have been employed in condensed matter and materials physics as models for the structures of a diverse range of substances from crystals and colloids to liquids, amorphous solids and glasses \cite{CLPCMP1995,HMTSL2006,ZallenPAS1983}. In structural biology, molecular dynamics simulations of interactions between large numbers of molecules employ chains of nonoverlapping spheres as models for various biological structures such as proteins and lipids \cite{DND2007a,Dokholyan2006a,DD2005a}.

In part due to the ability of these conceptually simple models to describe many of the fundamental characteristics of more complex substances, understanding the properties of sphere packings has also long been an area of interest in mathematics (for example, see \cite{CS1995a}). However, solving even some of the most basic of mathematical problems has proved challenging. For example, a proof of the Kepler conjecture, a proposition stating that the face-centered cubic lattice is the densest possible arrangement of spheres for $d=3$, has only recently emerged \cite{Hales2005a}. Furthermore, the kissing number $K_d$, or number of identical $d$-dimensional nonoverlapping spheres that can simultaneously be in contact with (kiss) a central sphere, was until recently only known rigorously for $d=1\!-\!3$, $8$ and $24$ \cite{CSSPLG1999}, though Musin \cite{Musin2008a} has now proved the $d=4$ case ($K_4 = 24$).

\begin{figure}[ht]
\centering
\includegraphics[width = 3.0in,viewport = 5 5 460
465,clip]{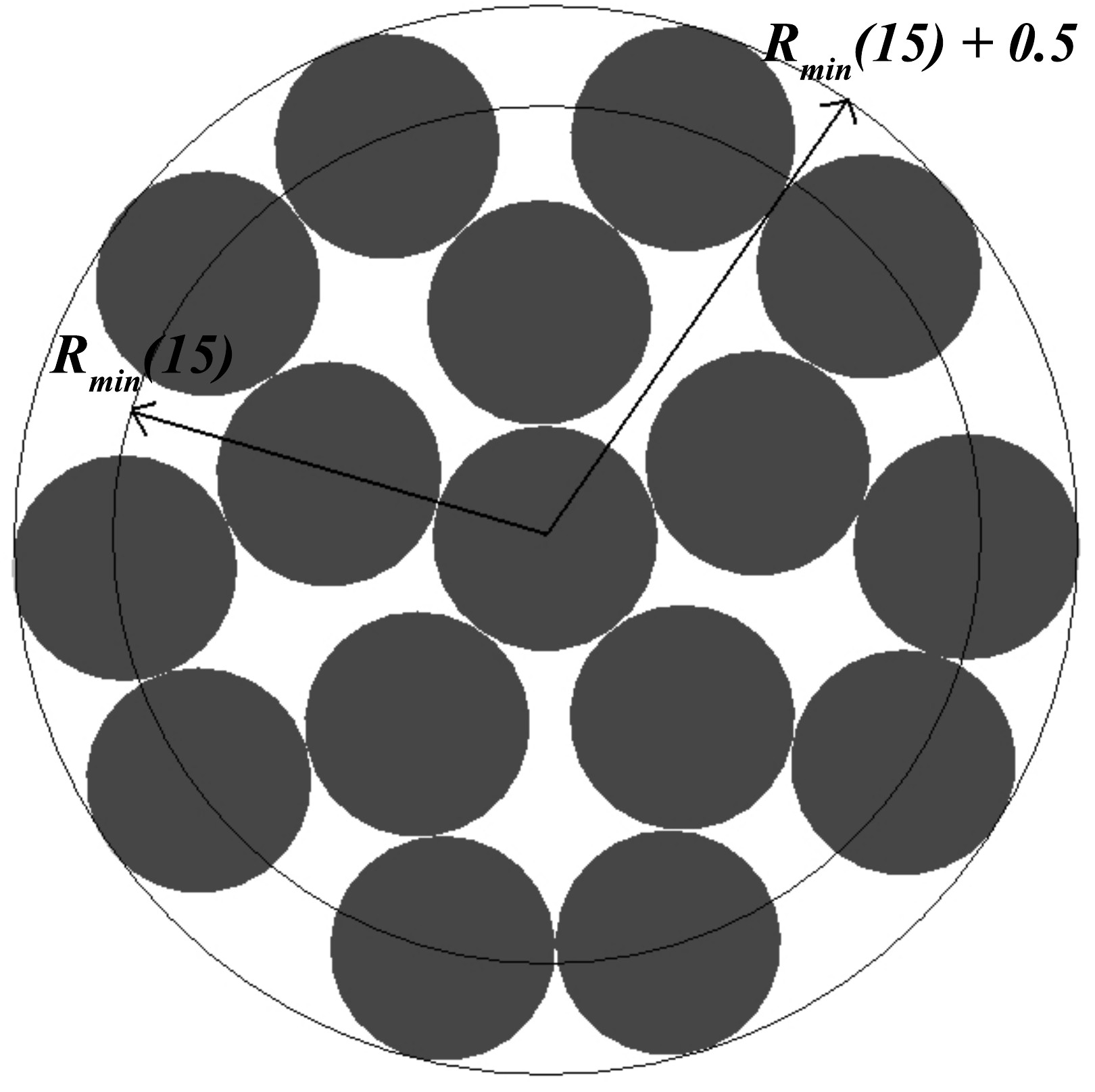}
\caption{A conjectured DLP optimal packing (point group $D_{5h}$) for $N=15$, $d=2$, $R_{min}(15) = 1.873123\dots$, with encompassing sphere of radius $R_{min}(15) + 0.5 = 2.373123\dots$.} 
\label{DLP15pic}
\end{figure}

One sphere packing problem that has not been generally addressed for an arbitrary number of spheres is that of finding the maximally dense (optimal) packing(s) of $N$ identical $d$-dimensional nonoverlapping spheres near (local to) an additional fixed central sphere such that the greatest radius $R$ from any of the surrounding spheres' centers to the center of the fixed sphere is minimized. This problem is called the densest local packing (DLP) problem \cite{HST2009b}. There is a single minimized greatest radius, denoted by $R_{min}(N)$, for each $N$ in the DLP problem in ${\mathbb R}^d$, though generally for each $N$ there may be multiple distinct packings that achieve this radius. Figure \ref{DLP15pic} depicts a conjectured optimal packing, belonging to point group $D_{5h}$ \cite{CottonCAGT1990}, for the DLP problem for $N=15$, $d=2$, with $R_{min}(15) = 1.873123\dots$ \cite{endnote3}.

In various limits, the densest local packing problem encompasses both the kissing number and (infinite) sphere packing problems. The former is a special case of the DLP problem in that $K_d$ is equal to the greatest $N$ for which $R_{min}(N) = 1$, and the latter is equivalent to the DLP problem in the limit that $N \rightarrow \infty$. The equivalence of the latter problem may be explained by observing that in the limit as $N \rightarrow \infty$, the boundary of radius $R_{min}(N) \rightarrow \infty$, and that in this limit the ratio of the number of spheres within a fixed finite distance of the boundary to the number in the bulk is zero.

The densest local packing problem is relevant to the realizability of functions that are candidates to be the pair correlation function of a packing of identical spheres. For a statistically homogeneous and isotropic packing, the pair correlation function is denoted $g_2(r)$; it is proportional to the probability density of finding a separation $r$ between any two sphere centers and normalized such that it takes the value of unity when no spatial correlations between centers are present. Specifically, no function can be the pair correlation function of a point process (where a packing of spheres of unit diameter is a point processes in which the minimum pair separation distance is unity) unless it meets certain necessary, but generally not sufficient, conditions known as realizability conditions \cite{Lenard1975a,TS2002a,KLS2007a}. Two of these conditions that appear to be particularly strong for the realizability of sphere packings \cite{TS2006a} are the nonnegativity of $g_2(r)$ and its corresponding structure factor $S(k)$, where
\begin{equation}
S(k) = 1 + \rho \tilde{h}(k)
\label{structFactCond}
\end{equation}
with number density $\rho$ and
\begin{equation}
\tilde{h}(k) = (2\pi)^{d/2}\int_0^{\infty}r^{d-1}h(r)\frac{J_{d/2-1}(kr)}{(kr)^{d/2-1}}dr
\label{dDimFourier}
\end{equation}
the $d$-dimensional Fourier transform of the total correlation function $h(k) \equiv g_2(r)-1$, with $J_{\nu}(x)$ the Bessel function of the first kind of order $\nu$.

The $g_2$-invariant process of Torquato and Stillinger \cite{TS2002a} is a method to maximize the number density $\rho$ associated with the structure factor $S(k)$ of a given parameterized family of test functions, where each function in the family is a candidate to be the pair correlation function of a statistically homogeneous and isotropic packing of spheres. In the $g_2$-invariant process, the problem of finding the maximal achievable $\rho$ is posed as an optimization problem: maximize $\rho$ over the parameters subject to the nonnegativity of the test function and its corresponding structure factor. This process could be improved by the addition of further realizability conditions on the pair correlation function, assuming that these further conditions included information beyond that incorporated in the two nonnegativity conditions discussed above.

Knowledge of the maximal number of sphere centers that may fit within radius $R$ from an additional fixed sphere center, where that maximal number is equal to the greatest $N$ in the DLP problem for which $R_{min}(N) \leq R$, may be employed to construct an additional realizability condition on $g_2(r)$. As was discussed in previous papers \cite{HST2009a,HST2009b}, this realizability condition has been shown to encode information not included in the nonnegativity conditions on pair correlation functions and their corresponding structure factors alone. 

The DLP problem may be alternatively stated as the problem of finding the densest packing of $N$ identical nonoverlapping spheres of unit diameter near an additional fixed sphere, where number density $\rho$ is measured over the volume enclosed by an \textit{encompassing sphere} of radius $R+0.5$ (see Fig. \ref{DLP15pic}) centered on the fixed sphere. We note that for $N$ spheres of unit diameter, the number density $\rho$ of the packing is linearly proportional (by a constant that varies only with dimension) to the packing fraction $\phi(R+0.5)$, the fraction of the volume of the encompassing sphere covered by the spheres of unit diameter. As will be discussed in detail later, the maximal infinite-volume packing fraction $\phi_*^{\infty}$ of identical nonoverlapping spheres in $d$ dimensions may be bounded from above by employing a specific definition of local packing fraction for a given number $N$ of spheres.

For small numbers of spheres ($N \leq 1000$) in low dimensions ($d \leq 10$), a algorithm combining a nonlinear programming method with a stochastic search of configuration space can be employed on a personal computer to find solutions to the DLP problem. Using such an algorithm, the details of which are outlined in Appendix \ref{algorithm}, we find and present putative DLP optimal packings and their corresponding $R_{min}(N)$ in ${\mathbb R}^2$ for $N = 1$ to $N=109$, and for the values of $N$ corresponding to full shells of the triangular lattice from $N=120$ to $N=348$. Though we recognize that the putative optimal packings found by our algorithm are not rigorously proved to be optimal, we analyze each configuration of $N$ spheres under the assumption that it is a global minimum of the DLP problem. This assumption of optimality is supported by the proved robustness of the algorithm in recovering the known and strongly conjectured global minima of the DLP problem (e.g., the kissing numbers for $d=1\!-\!4$ and for $d=8$, the curved hexagonal packings for $d=2$; $N=18$, $36$, $60$, $90$ and $126$ \cite{LG1997a}) and by repeated testing.

The aforementioned realizability condition on the pair correlation function is valid whether or not the putative optimal packings we have found are indeed global minima, as global minima simply provide the most restrictive realizability condition. However, the upper bound on the maximal density of an infinite sphere packing requires knowledge of proved optimal $R_{min}(N)$ to be rigorously correct, though we have found in practice that for $d=2$ and over the range of $N$ tested that our putative bound is valid. With regard to this finding and the proved robustness of the algorithm over the range of $N$ studied, in the following sections we refer to all DLP packings and $R_{min}(N)$ presented as optimal.

In Sec. II, we discuss the realizability condition that results from knowledge of a finite number of $R_{min}(N)$ in a space ${\mathbb R}^d$ of arbitrary dimension $d$. We present the condition derived from knowledge of $R_{min}(N)$ for $N=1$ to $N=109$ in ${\mathbb R}^2$ and compare the $R_{min}(N)$ values to the shell distances in a triangular lattice of $N$ disks. In Sec. III, we construct a logical argument to prove the validity of the aforementioned upper bound, and we present the $d=2$ upper bounds derived from our method for selected $N$ from $N=6$ to $N=348$. In Sec IV, we present $d=2$ optimal packings and their corresponding $R_{min}(N)$ for selected values of $N$ from $N=10$ to $N=348$. We analyze the optimal packings presented and discuss their symmetry characteristics, noting that there is significant variability in both the configurations and symmetry elements of optimal packings over the range of $N$ studied. In Sec. V, we summarize our results and findings, and we discuss some of the implications of our work.

In a sequel to this paper, we will present and analyze DLP optimal packings and their corresponding $R_{min}(N)$ for the $d=3$ case over a similar range of $N$. We will compare $R_{min}(N)$ values to shell distances in Barlow packings \cite{Barlow1883a}, where the Barlow packings, of which the best-known are face centered cubic (FCC) and hexagonal close packed (HCP) arrangements, all individually achieve the maximal infinite-volume packing fraction for $d=3$, $\phi_*^{\infty} = \pi/\sqrt{18} = 0.740481\dots$.

The coordinates for and images of DLP optimal packings and values for $R_{min}(N)$ over the entire range of $N$ studied can be found on the authors' website \cite{website}. 

\section{Pair correlation function realizability and $Z_{max}(R)$}

The realizability condition on $g_2(r)$ results from a relation between an upper bound on the maximal value of the function $Z(R)$, to be defined shortly, and $g_2(r)$. The function $Z({\bf r}_i,R)$ is defined for packings of nonoverlapping spheres of unit diameter as the number of sphere centers that are within distance $R$ from a (additional) sphere center at position ${\bf r}_i$, with $i$ an index over centers. The maximum over all ${\bf r}_i$ of $Z({\bf r}_i,R)$ is an upper bound on the maximum of the (different) function $Z(R)$, where $Z(R)$ is defined for a statistically homogeneous packing as the expected number of sphere centers within distance $R$ from any given sphere center, or equivalently as the average of $Z({\bf r}_i,R)$ over all $i$. The function $Z(R)$ can be related to the pair correlation function $g_2(r)$, where for a pair correlation function $g_2({\bf r})$ that is direction-dependent, $g_2(r)$ is the directional average of $g_2({\bf r})$, by
\begin{equation}
Z(R) = \rho s_1(1)\int_0^Rx^{d-1}g_2(x)dx.
\label{ZR}
\end{equation}
In Eq. (\ref{ZR}), $\rho$ is the constant number density of sphere centers and $s_1(r)$ is the surface area of a sphere of radius $r$ in ${\mathbb R}^d$,
\begin{equation}
s_1(r) = \frac{2\pi^{d/2}r^{d-1}}{\Gamma(d/2)}.
\label{s1}
\end{equation}

The maximum at fixed $R$ of the function $Z({\bf r}_i,R)$ over all possible configurations of sphere centers $\{{\bf r}_i\}$ is equal to the greatest number $N$ for which $R_{min}(N) \leq R$ for a DLP optimal packing of $N$ spheres in ${\mathbb R}^d$. Defining $Z_{max}(R)$ for all $R$ as this greatest $N$, it follows that
\begin{equation}
Z(R) \leq Z_{max}(R)
\label{ZRbound}
\end{equation}
for any sphere packing.

\begin{figure}[ht]
\centering
\includegraphics[width = 5.5in,viewport = 20 25 735
540,clip]{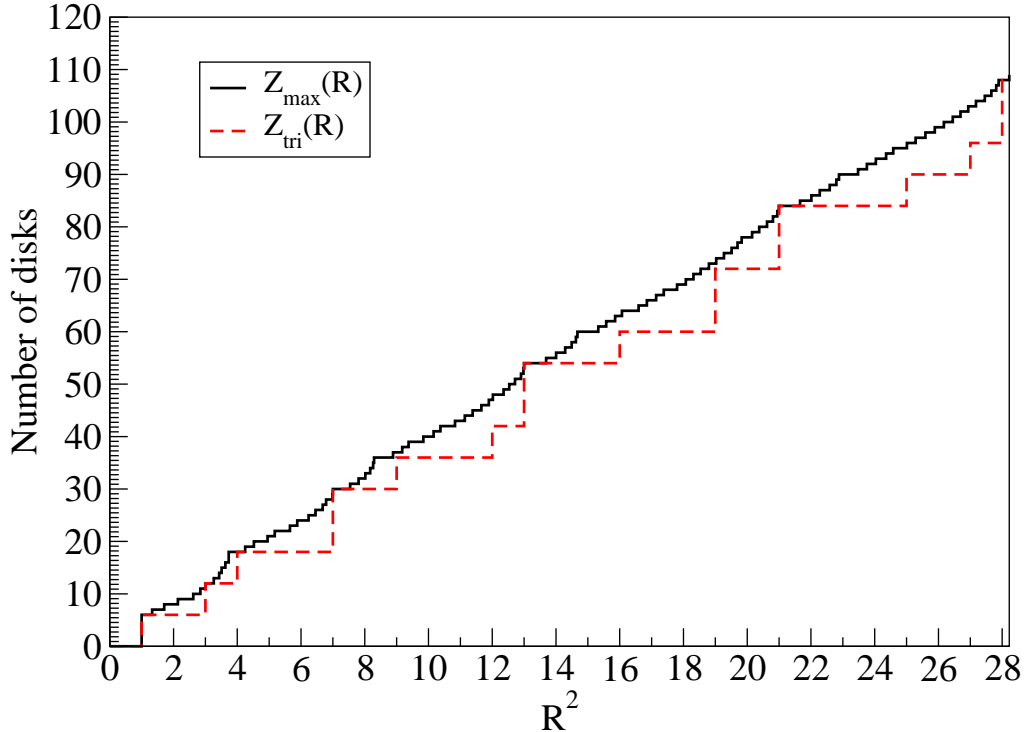}
\caption{$Z_{max}(R)$ vs $R^2$, as determined by optimal and putative optimal solutions to the DLP problem for $N=1$ to $N=109$, and $Z_{tri}(R)$. The radius $R$ of the (larger) disk enclosing the centers of the $N$ (smaller) disks and fixed disk is measured in units of the diameter of the enclosed disks.} 
\label{ZmaxVStri}
\end{figure}

Equation (\ref{ZRbound}) is a realizability condition on $g_2(r)$, with $Z(R)$ in ${\mathbb R}^d$ defined in terms of $g_2(r)$ in Eq. (\ref{ZR}) and $Z_{max}(R)$ defined completely by the solutions to the DLP problem over all $N$. In ${\mathbb R}^2$, the function $Z_{max}(R)$ may be compared to the function $Z_{tri}(R)$, with $Z_{tri}(R)$ defined as the sum of the number of disk centers included in all (full) shells of radius less than or equal to $R$ in a triangular lattice of contacting disks. Both $Z_{max}(R)$ and $Z_{tri}(R)$ increase roughly linearly with $R^2$, as the area of a disk of radius $R$ is proportional to $R^2$. Clearly, $Z_{max}(R) \geq Z_{tri}(R)$ for all $R$, as can be seen in Fig. \ref{ZmaxVStri}, a plot of $Z_{max}(R)$ vs. $R^2$ for $N=1\!-\!109$ and $d=2$ alongside a plot of $Z_{tri}(R)$. 

In arbitrary dimension $d$, the function $Z_{max}(R)$ is zero for $R < 1$ due to the nonoverlap condition. For $R=1$, $Z_{max}(R)$ in ${\mathbb R}^d$ is equal to the kissing number $K_d$. For $R > 1$, $Z_{max}(R)$ should grow approximately as $R^d$ in proportion with the growth of the volume of a $d$-dimensional sphere, though in a separate work \cite{HST2009b} we have proved that this cannot be the case for $R \leq \tau$, with $\tau = (1+\sqrt(5))/2$ the golden ratio. For $R \leq \tau$, $Z_{max}(R)$ in any dimension cannot exceed the maximal number of sphere centers that can be placed on the surface of a sphere of radius $R$. Alternatively stated, this counterintuitive result requires that for $R \leq \tau$, $Z_{max}(R)$ can grow only as the surface area $R^{d-1}$. Specifically for $d=2$, $3$ and $4$, $Z_{max}(R\leq\tau)$ is less than or equal to $10$, $33$ and $120$, respectively. 

\section{Bounds on infinite sphere packings and the DLP problem}

We discuss two distinct methods through which the function $Z_{max}(R)$ in ${\mathbb R}^d$ can be employed to bound from above the maximal infinite-volume packing fraction $\phi_*^{\infty}$ of an infinite packing of identical nonoverlapping spheres. The first has been discussed in detail in two separate works \cite{HST2009a,HST2009b}; it is precisely the method of Cohn and Elkies in \cite{CE2003a}. In Ref. \cite{CE2003a}, the authors employ an infinite-dimensional linear program that is the dual of the $g_2$-invariant program \cite{TS2006a} discussed in Sec. I to find the best known bounds on the maximal infinite-volume packing fraction $\phi_*^{\infty}$ of sphere packings at least in dimensions four through 36. An improved method to bound $\phi_*^{\infty}$ from above adds the information encoded in the $Z_{max}(R)$ realizability condition to augment the approach of Cohn and Elkies in \cite{CE2003a} as proposed by Cohn, Kumar and Torquato \cite{CKT2009a}.

The second method bounds $\phi_*^{\infty}$ from above by the maximal local packing fraction $\hat{\phi}_*(N)$ of a packing of a number $N$ of identical nonoverlapping spheres around an additional central sphere. The local packing fraction $\hat{\phi}(N)$, of which $\hat{\phi}_*(N)$ is the maximum, is defined for $N$ spheres around an additional fixed central sphere as the total volume of the $N+1$ spheres divided by the volume of a sphere of radius $R$, where $R$ is, as in the DLP problem, the greatest of the distances from the centers of the $N$ surrounding spheres to the center of the fixed sphere. From this definition of $\hat{\phi}(N)$, the maximal local packing fraction $\hat{\phi}_*(N)$ for $N$ $d$-dimensional spheres of unit diameter takes the form
\begin{equation}
\hat{\phi}_*(N) = \frac{N+1}{(2R_{min}(N))^d},
\label{maxLocalDensity}
\end{equation}
where $R_{min}(N)$ is as before the optimal radius in the DLP problem for $N$ spheres in ${\mathbb R}^d$. For the sake of convenience, we have collected and defined the various packing fraction terms used in this section in table \ref{packTable}.

\begin{table}
\caption{Packing fraction terms and definitions used in the text.} \label{packTable} 
\begin{center}
\begin{tabular}{|c c l|}
\hline
{\bf Symbol} & {\bf Term} & {\bf Definition} \\
\hline
\hline
\multirow{2}{*}{$\phi(R)$} & \multirow{2}{*}{\,\,\,{\bf packing fraction}\,\,\,} & volume fraction of a (larger) sphere of radius $R$ covered \\ & & by identical nonoverlapping spheres of unit diameter \\
\hline
\multirow{2}{*}{$\phi_{max}(R)$} & {\bf maximal} & \multirow{2}{*}{greatest achievable $\phi(R)$ for a given $R$} \\ & {\bf packing fraction} & \\
\hline
\multirow{2}{*}{$\phi^{\infty}$} & {\bf infinite-volume} & fraction of space covered by identical nonoverlapping \\ &  {\bf packing fraction} & spheres in a given infinite packing \\
\hline
\multirow{3}{*}{$\phi_*^{\infty}$} & {\bf maximal} & \\ & {\bf infinite-volume} & greatest achievable infinite-volume packing fraction \\ & {\bf packing fraction} & \\
\hline
\multirow{4}{*}{$\hat{\phi}(N)$} & & ratio of the sum of the volumes of a nonoverlapping fixed \\ & {\bf local} & central sphere of unit diameter and its surrounding $N$ \\ & {\bf packing fraction} & same-size spheres to the volume of a (larger) sphere of \\ & & radius $R$, with $R$ defined as in the DLP problem \\
\hline
\multirow{2}{*}{$\hat{\phi}_*(N)$} & {\bf maximal local} & \multirow{2}{*}{greatest achievable $\hat{\phi}(N)$ for a given $N$ in ${\mathbb R}^d$; see Eq. (\ref{maxLocalDensity})\,\,} \\ & {\bf packing fraction} & \\
\hline
\multirow{2}{*}{$\bar{\phi}(R)$} & {\bf average local} & average packing fraction within a window of radius $R$ of \\ & {\bf packing fraction} & identical nonoverlapping spheres of unit diameter \\
\hline 
\end{tabular}
\end{center}
\end{table}

The statement that $\hat{\phi}_*(N)$ bounds from above $\phi_*^{\infty}$ for certain $N$ relies on a construction that links local packing fraction $\hat{\phi}(N)$ to the infinite-volume packing fraction $\phi^{\infty}$ of a packing of identical nonoverlapping spheres. The construction proceeds as follows. First, a spherical window of radius $R_{min}(N)$ is centered on an arbitrary sphere in a single configuration of an infinite packing of identical nonoverlapping spheres of unit diameter. An infinite packing of identical nonoverlapping spherical windows is created by replicating the initial window infinitely many times and placing the (replicated) window centers in the exact (scaled) spatial configuration of the centers of the original infinite packing of spheres of unit diameter. The only difference between the two configurations is that the configuration of windows is scaled by $2R_{min}(N)$, the ratio of the radius of a window to the radius of a sphere of unit diameter. 

\begin{figure}[ht]
\centering
\includegraphics[width = 5.8in,viewport = 0 120 815
425,clip]{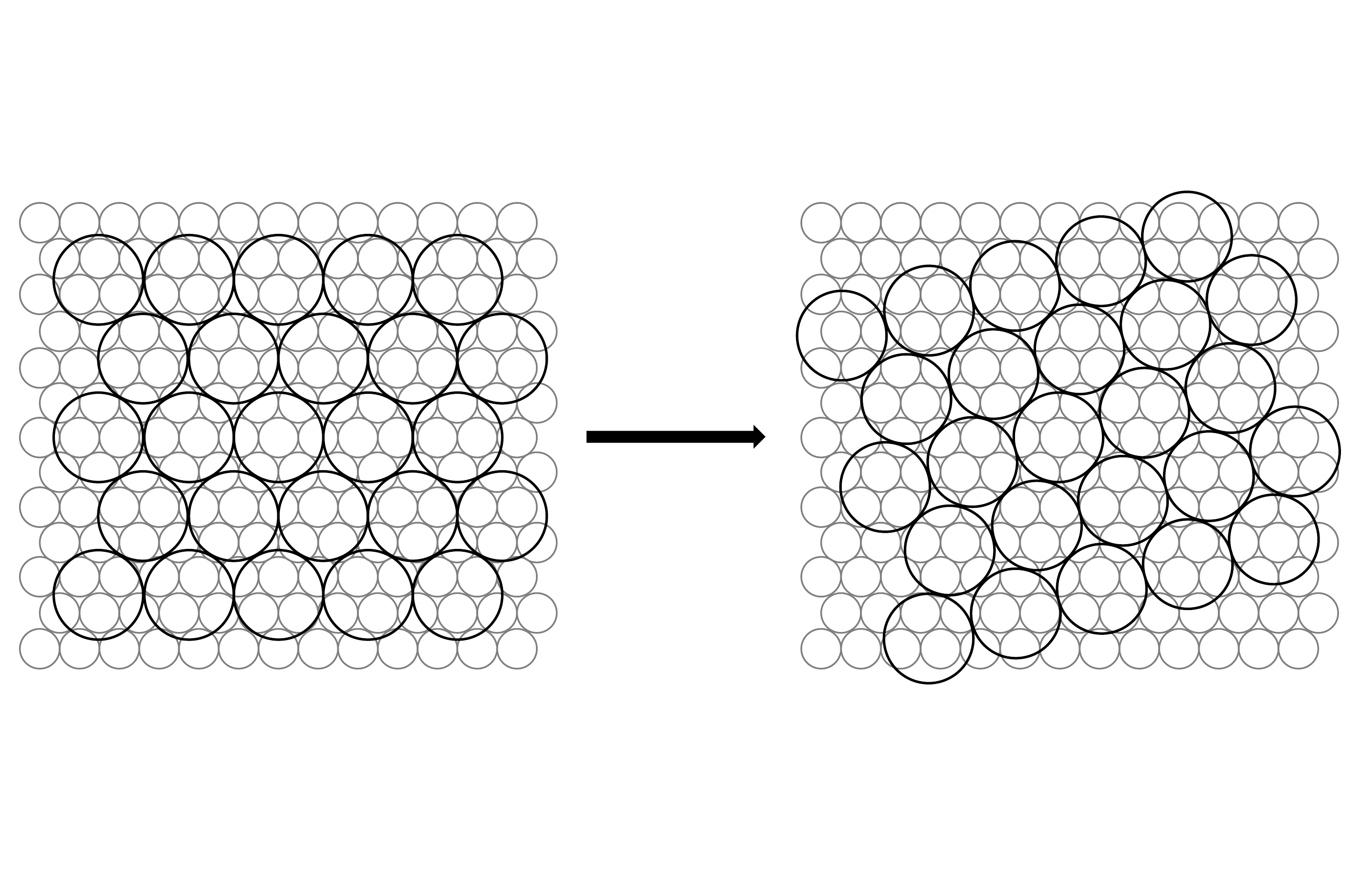}
\caption{Illustration of a rotation of an infinite packing of identical nonoverlapping windows of radius $R$, arranged on the sites of a triangular lattice, overlayed upon an infinite packing of smaller identical nonoverlapping disks. A rotation is selected such that irrational ratios are achieved between the components of at least one of the lattice vectors of the packing of windows in the directions of the lattice vectors of the packing of smaller disks. As a result, at large distances from the axis of rotation, any window can be thought of as being placed at random onto the packing of smaller disks. It follows that the average local packing fraction $\bar{\phi}(R)$ of the smaller disks within the windows is equal to the infinite-volume packing fraction $\phi^{\infty}$ of the smaller disks.}
\label{spheresWindows}
\end{figure}

As will be made precise in the following paragraphs, for any such packing of windows and spheres of unit diameter, a rigid rotation for the overlayed packing of windows can be found such that the average local packing fraction $\bar{\phi}(R_{min}(N))$ of spheres of unit diameter within the windows is equal to the infinite-volume packing fraction $\phi^{\infty}$ of the spheres of unit diameter in ${\mathbb R}^d$. The concepts of overlay and rotation are illustrated in Fig. \ref{spheresWindows} for a triangular lattice of disks of unit diameter. 

It suffices to apply the aforementioned construction to periodic packings, as it has been shown that periodic packings in ${\mathbb R}^d$ can obtain an infinite-volume packing fraction $\phi^{\infty}$ arbitrarily close to the maximal infinite-volume packing fraction $\phi_*^{\infty}$ (for example, see \cite{CE2003a}). A periodic packing can be defined in terms of a lattice $\Lambda$, where $\Lambda$ in ${\mathbb R}^d$ is a subgroup consisting of the integer linear combinations of a set of vectors that constitute a basis for ${\mathbb R}^d$. For identical nonoverlapping spheres, a lattice packing is a packing where the centers of the spheres are located at the points of $\Lambda$. In such a lattice packing, the space ${\mathbb R}^d$ can be divided into finite-size identical nonoverlapping regions called fundamental cells, each containing the center of only one sphere.

A periodic packing is a more general formulation of a lattice packing. For identical nonoverlapping spheres, a periodic packing is obtained by placing a fixed configuration of a number $M$ of spheres in a fundamental cell that is then periodically replicated (without overlap between cells or spheres) to cover ${\mathbb R}^d$. The fixed configuration of $M$ spheres within each fundamental cell is arbitrary subject only to the overall nonoverlap condition of the periodic packing of spheres. As used here, the term ``lattice'' is the same as ``Bravais lattice'' conventionally used in the physics literature.

Consider an infinite periodic packing (with lattice basis vectors $\{{\bf u}_j\}$) of nonoverlapping spheres of unit diameter in ${\mathbb R}^d$, $d > 1$, with infinite-volume packing fraction $\phi^{\infty}$. Place an infinite periodic packing (with lattice basis vectors $\{{\bf v}_j\}$) of identical nonoverlapping windows of radius $R_{min}(N)$ over the infinite periodic packing of spheres of unit diameter in the manner of the construction discussed previously. For the radius $R_{min}(N)$ of the windows, any positive integer $N_* \in {\mathbb N}$ can be considered, where in ${\mathbb R}^d$ the set ${\mathbb N}$ is defined such that each $N_*$ is the greatest number $N$ of spheres for which $R_{min}(N_*) = R_{min}(N)$. For example, with $N = 1$ in two dimensions, $R_{min}(1) = 1$, and the greatest $N$ for which $R_{min}(N) = 1$ is $N=N_*=6$ \cite{endnote5}. It is intuitively clear that for any greatest such $N = N_*$ and $R = R_{min}(N_*)$ that 
\begin{equation}
\hat{\phi}_*(N_*) \geq \phi_{max}(R_{min}(N_*)),
\label{phiStarVsPhiMax}
\end{equation}
where $\phi_{max}(R)$ is defined as the maximal fraction of space that identical nonoverlapping spheres of unit diameter may cover in a spherical window of radius $R$ \cite{endnote6}.

\begin{figure}[ht]
\centering
\includegraphics[width = 5.8in,viewport = 5 190 715
430,clip]{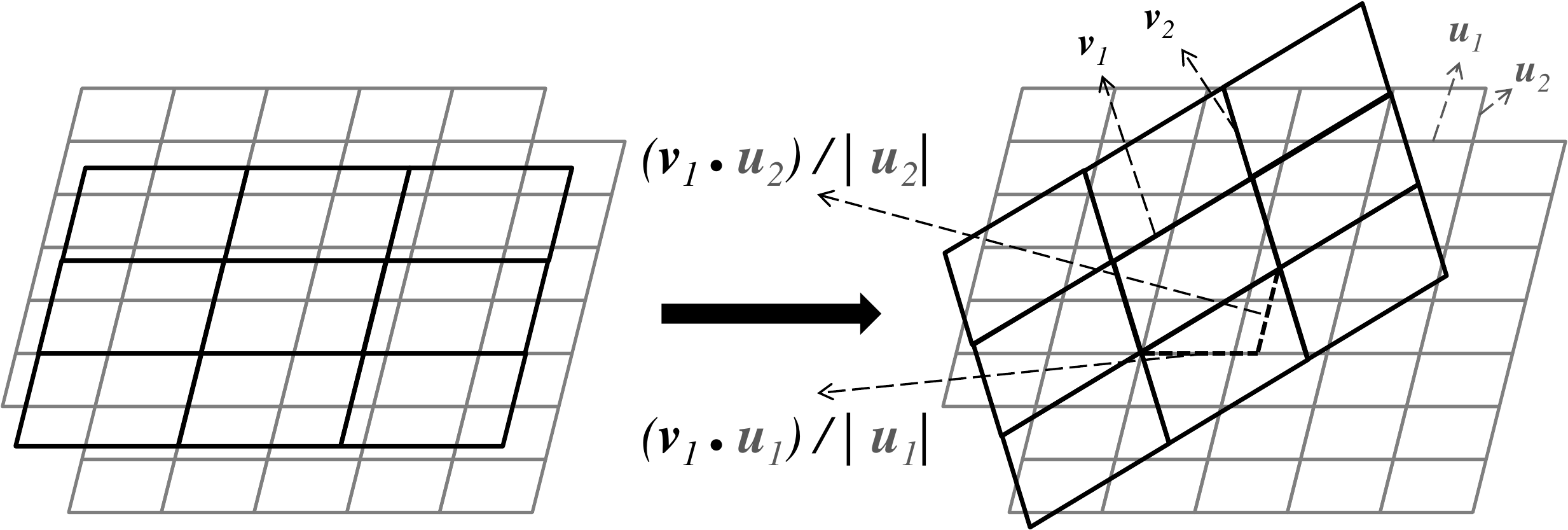}
\caption{Illustration of a rigid rotation of a window packing with parallelogram fundamental cells such that the ratios of the components of ${\bf v}_1$ in the directions of each of ${\bf u}_1$ and ${\bf u}_2$ to the magnitudes of ${\bf u}_1$ and ${\bf u}_2$, respectively, are both irrational.}
\label{cellsWindows}
\end{figure}

Now rigidly rotate the packing of windows around the center of the original window as in Fig. \ref{spheresWindows}. The component of a window-packing basis vector ${\bf v}_n$ that is in the direction of a given unit-diameter sphere packing basis vector ${\bf u}_m$ is ${\bf v}_n \cdot {\bf u}_m/|{\bf u}_m|$. A rotation may be found such that the ratios $\{{\bf v}_n \cdot {\bf u}_j/|{\bf u}_j|^2\}$ of the components of (at least) one of the window basis vectors ${\bf v}_n$ in the directions of each of the sphere packing basis vectors $\{{\bf u}_j\}$ to the respective magnitudes $\{|{\bf u}_j|\}$ of the sphere packing basis vectors are all irrational. This concept is illustrated in Fig. \ref{cellsWindows} in ${\mathbb R}^2$ for parallelogram fundamental cells.

After the rotation, due to the irrationality of the ratios of lattice vector components, the average fraction of space $\bar{\phi}(R_{min}(N))$ covered by the spheres of unit diameter in each window of the window packing is equal to the infinite-volume packing fraction $\phi^{\infty}$ of the unit-diameter spheres.

\begin{figure}[ht]
\centering
\includegraphics[width = 5.5in,viewport = 20 35 765
525,clip]{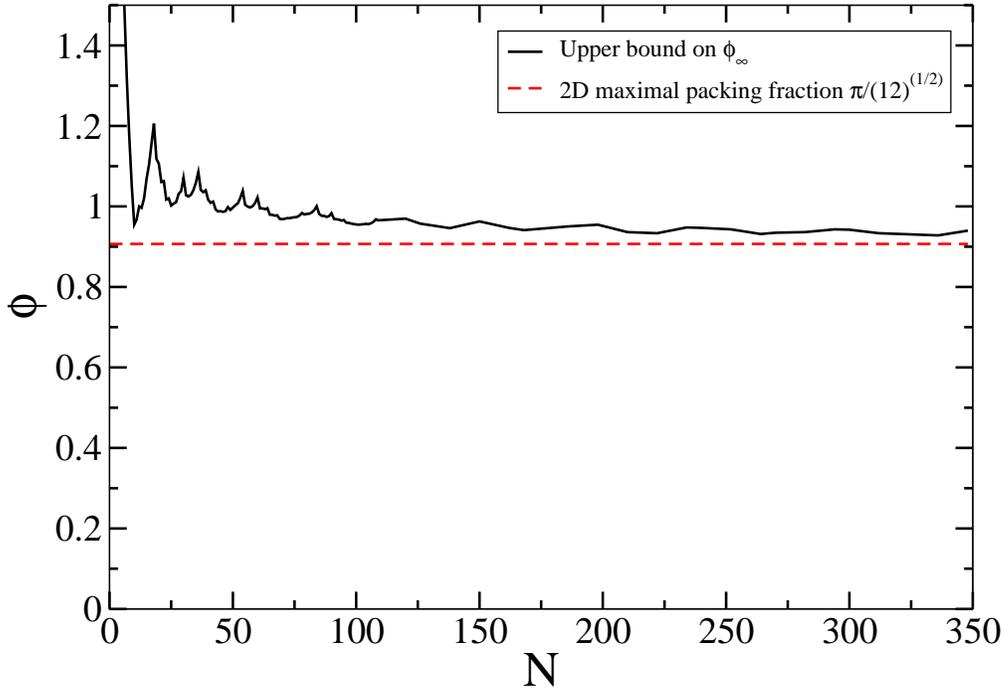}
\caption{maximal packing fraction $\phi_*^{\infty} = 0.906900\dots$ compared to putative upper bounds on $\phi_*^{\infty}$ for $d=2$ as determined by putative optimal solutions to the DLP problem for selected $N_* \in {\mathbb N}$ up to $N_*=348$. The minimum upper bound determined here is at $N_*=336$, $\hat{\phi}_*(336) = 0.928114$.} 
\label{upperBoundPlot}
\end{figure}

With this construction, the packing fraction $\phi^{\infty}$ of the spheres of unit diameter, equal to the average local packing fraction $\bar{\phi}(R_{min}(N))$ of the unit-diameter spheres within a window, can be compared to the maximal local packing fraction $\hat{\phi}_*(N)$ of that same window. As $\phi_{max}(R_{min}(N))$ is clearly greater than or equal to $\phi^{\infty}$, using Eq. (\ref{phiStarVsPhiMax}) we have
\begin{equation}
\hat{\phi}_*(N_*) \geq \phi_{max}(R_{min}(N_*)) \geq \phi^{\infty},
\label{upperBoundInt}
\end{equation}
for $N_* \in {\mathbb N}$ in ${\mathbb R}^d$.

Equation (\ref{upperBoundInt}) is true for all feasible $\phi^{\infty}$ including the maximal infinite-volume packing fraction $\phi_*^{\infty}$ of nonoverlapping spheres in ${\mathbb R}^d$. The maximal local packing fraction $\hat{\phi}_*(N_*)$ is therefore an upper bound on the maximal infinite-volume packing fraction $\phi_*^{\infty}$, or
\begin{equation}
\hat{\phi}_*(N_*) \geq \phi_*^{\infty}, \qquad N_* \in {\mathbb N}.
\label{upperBound}
\end{equation}
Figure \ref{upperBoundPlot} plots for $d=2$ and for selected values of $N_* \in {\mathbb N}$ up to $N_*=348$ the putative upper bound defined by $\hat{\phi}_*(N_*)$ alongside the proved maximal packing fraction, $\phi_*^{\infty} = \pi / \sqrt{12}$.

While the maximal infinite-volume packing fractions for $d$-dimensional identical nonoverlapping spheres are known with analytical rigor for $d=1\!-\!3$, $d=8$ and $d=24$ \cite{CK2010a}, this method could be used to improve upon the upper bounds on maximal packing fraction $\phi_*^{\infty}$ in dimensions where a value for $\phi_*^{\infty}$ has not not yet been proved. It is important to reiterate though that to be rigorous, the upper bound as defined requires a DLP radius $R_{min}(N)$ that is proved optimal. 

\section{Optimal packings in two dimensions}

The packing of nonoverlapping disks that uniquely achieves the highest infinite-volume packing fraction $\phi_*^{\infty} = \pi/\sqrt{12}$ for $d=2$ is well-known to be the packing such that each disk is in contact with exactly six others, with centers arranged on the sites of the triangular lattice. As DLP optimal packings are packing fraction-maximizing arrangements of $N$ disks centered on a (additional) disk, one might expect that all or a major subset of disks in any given DLP optimal packing will always sit on the sites of the triangular lattice.

Over the range of $N$ studied, however, this is in fact very infrequently the case. We find that at only three values of $N$ greater than $6$, at $N = 12$, $30$ and $54$, are triangular lattice configurations also DLP optimal configurations, while most DLP optimal packings are significantly more locally dense than a packing of $N$ disks around a central disk with centers on the sites of the triangular lattice, as is illustrated in Fig. \ref{ZmaxVStri} for $N=1\!-\!109$.

In general, we find wide variation in the symmetries and other characteristics of DLP optimal packings. For the majority of cases, there appear to be an uncountably infinite number (a continuum) of optimal configurations of $N$ sphere centers at optimal $R_{min}(N)$, with the continuum attributable to the presence of rattlers. A rattler in a packing of spheres in ${\mathbb R}^d$ is a sphere that is positioned such that it may be individually moved in at least one direction without resulting overlap of any other sphere within the packing or the packing boundary (in this case, the encompassing sphere of radius $R_{min}(N)+0.5$), i.e., a rattler is a sphere that is not locally jammed \cite{TTD2000a,TS2001a}. The rattlers present in the following figures are indicated by a lighter shading, while the fixed central spheres (disks) are indicated by an open circle.

\subsection{Packings that are proved optimal}
In two dimensions for $N \leq 10$, we proved \cite{HST2009b} that optimal packings are those in which $R_{min}(N)$ is equal to the radius of the minimal-radius circle onto which the centers of $N$ disks may be packed. These $R_{min}(N)$ may be analytically calculated via simple trigonometry, yielding $R_{min}(N) = 1$ for $N \leq 6$, and
\begin{equation}
R_{min}(N) = \frac{1}{\sqrt{2(1-\cos{2\pi/N})}}, \qquad 6 \leq N \leq 10.
\label{Rmin10}
\end{equation}

\begin{figure}[!ht]
\centering
\subfigure[]{\includegraphics[width=2.0in]{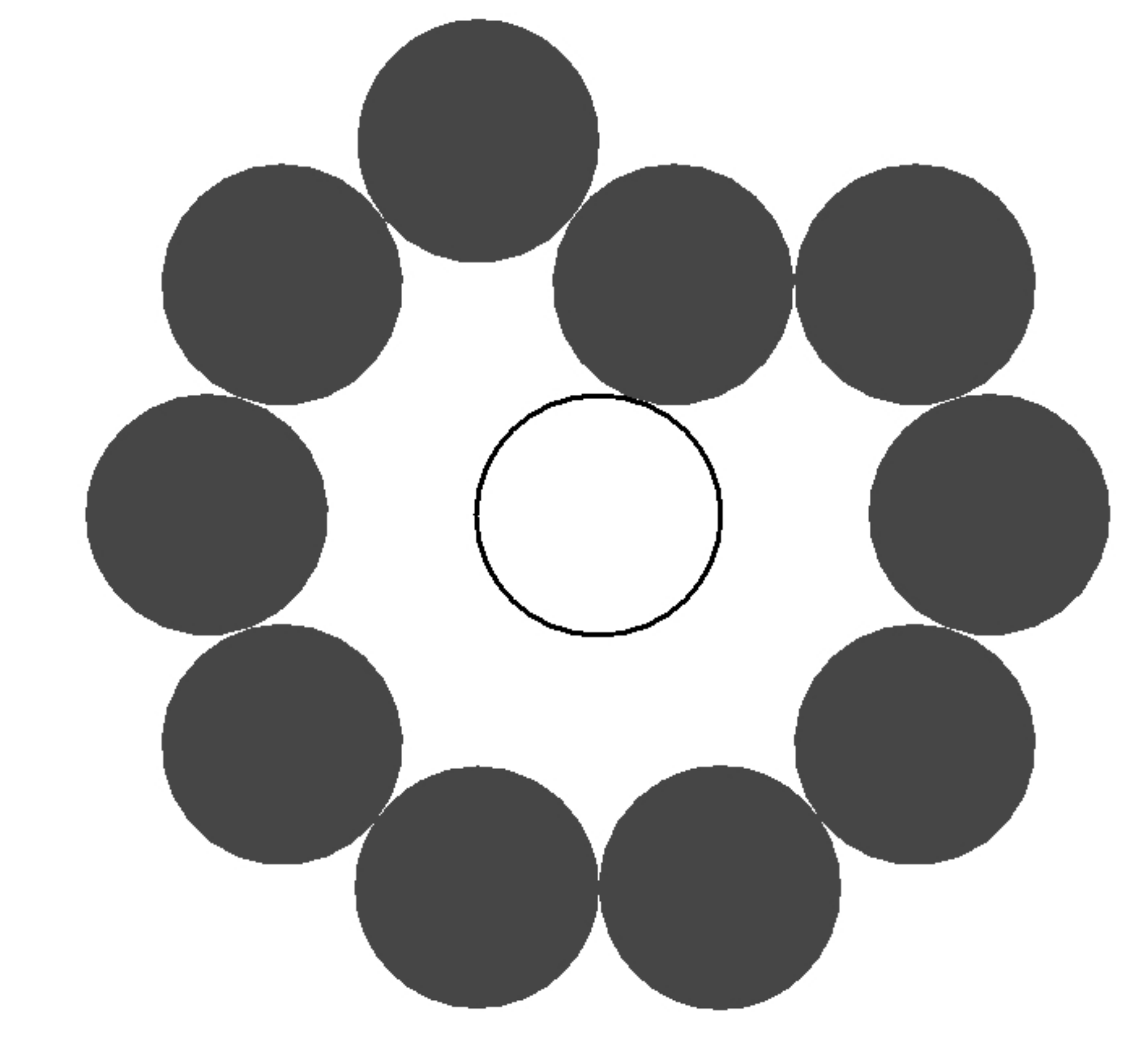}} \\
\subfigure[]{\includegraphics[width=2.0in]{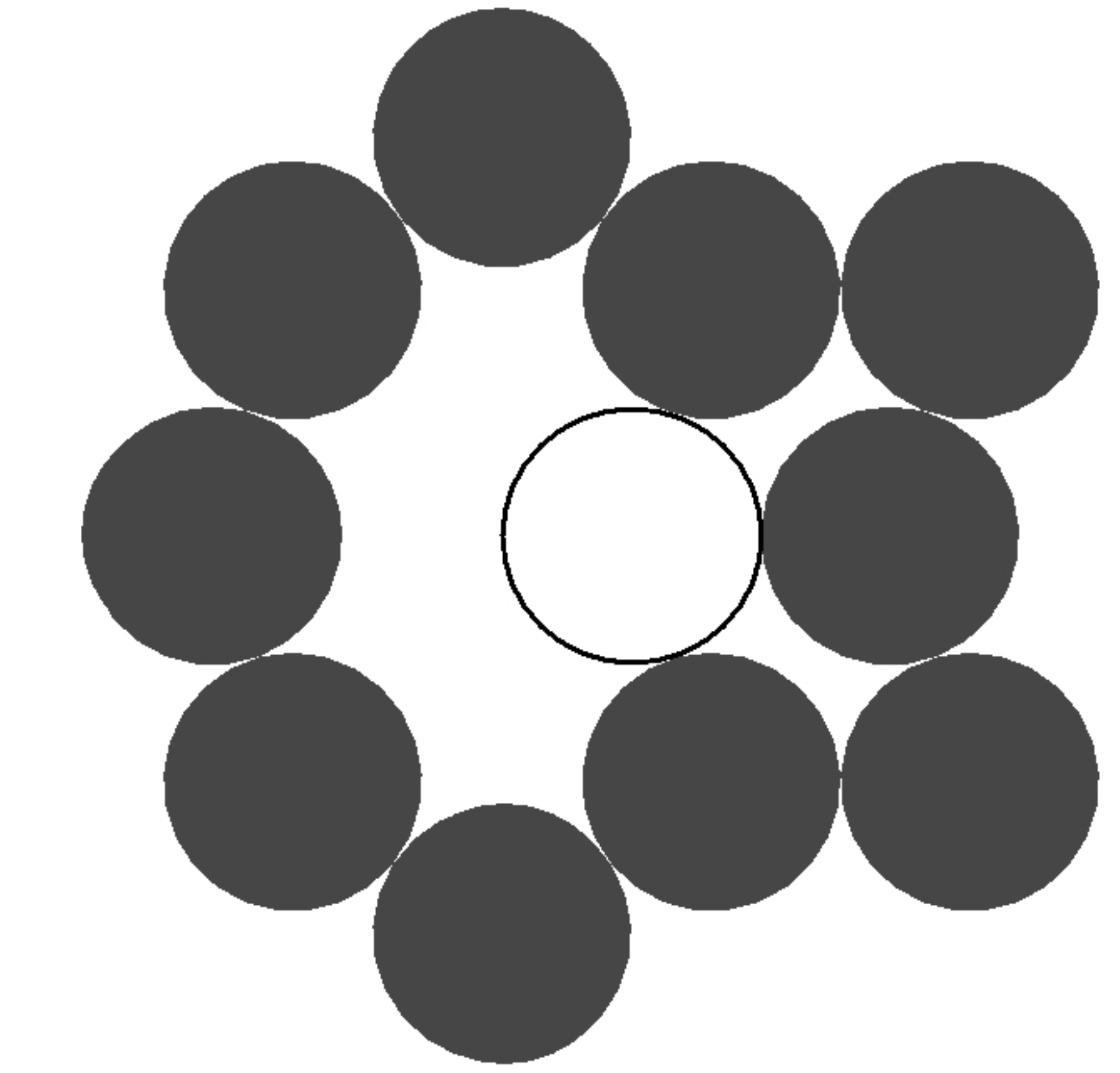}} \\
\subfigure[]{\includegraphics[width=2.0in]{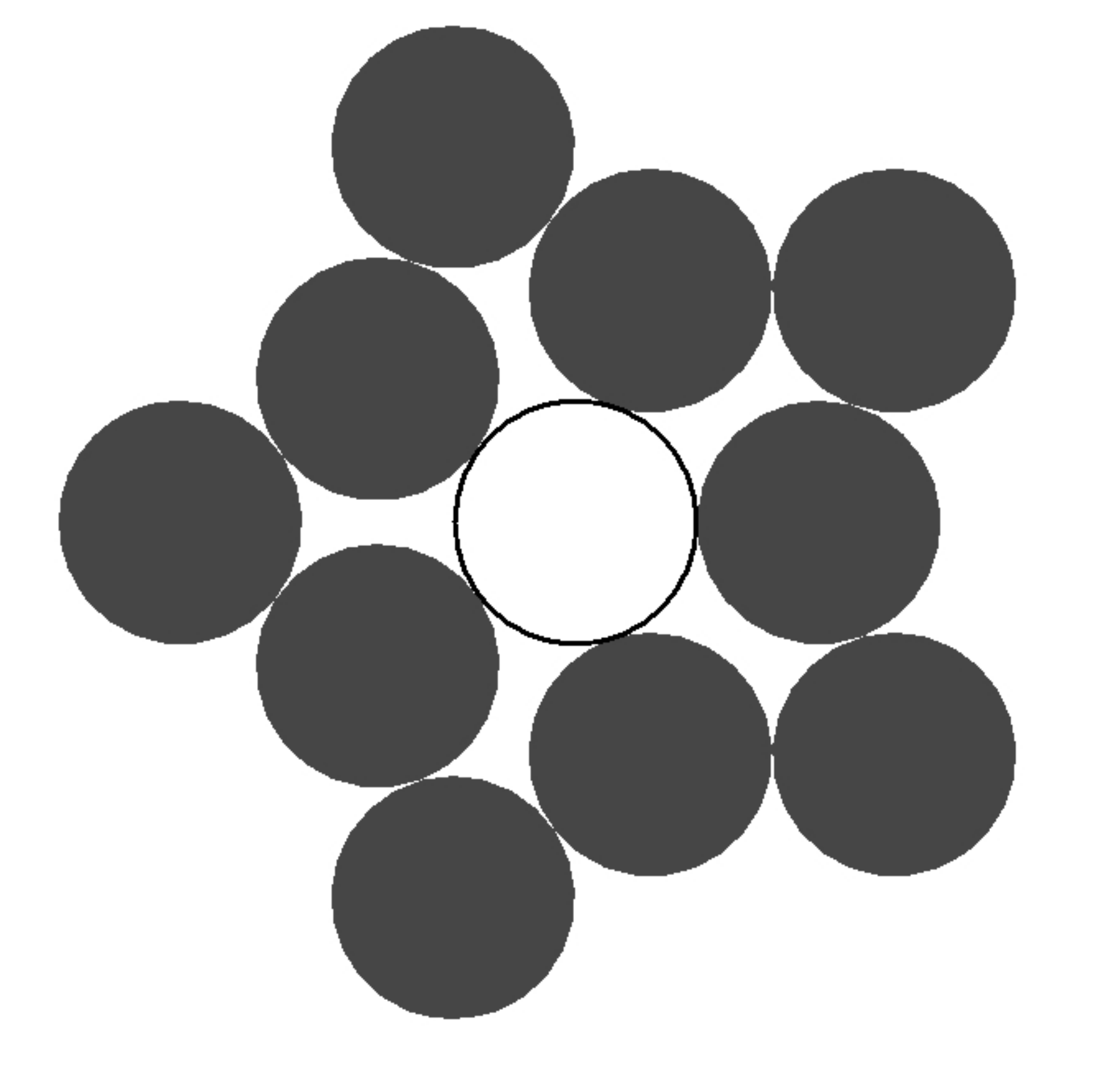}}
\caption{Three of a finite number of optimal cases for $N=10$, $R_{min}(10) = \tau = 1.618204\dots$ formed by radially translating disks whose centers lie on a circle of radius $R_{min}(10) = \tau$ to be in contact with the fixed central disk. (a) point group $C_{2v}$. (b) point group $C_{2v}$. (c) point group $D_{5h}$.}
\label{10opt}
\end{figure}

For $6 \leq N \leq 9$, DLP optimal packings with $R_{min}(N)$ as defined in Eq. (\ref{Rmin10}) are unique up to rotations and correspond to configurations where all sphere centers lie on the circle of radius $R_{min}(N)$ at distance unity from two adjacent centers. In general, for many $N$ there is a unique (up to rotations) optimal packing, and in a smaller number of cases such as for $N=10$, $60$, $90$ and $126$, there are a finite number of degenerate optimal packings. Figure \ref{10opt} depicts three of a finite number of optimal packings for $N=10$, $d=2$, where $R_{min}(10) = \tau$; these packings are formed by radially translating up to five disks with centers on the circle at radius $R_{min}(10) = \tau$, the golden ratio, to be in contact with the fixed central disk at distance unity.

\subsection{Curved hexagonal packings}
\begin{figure}[!ht]
\centering
\subfigure[]{\includegraphics[width=2.0in]{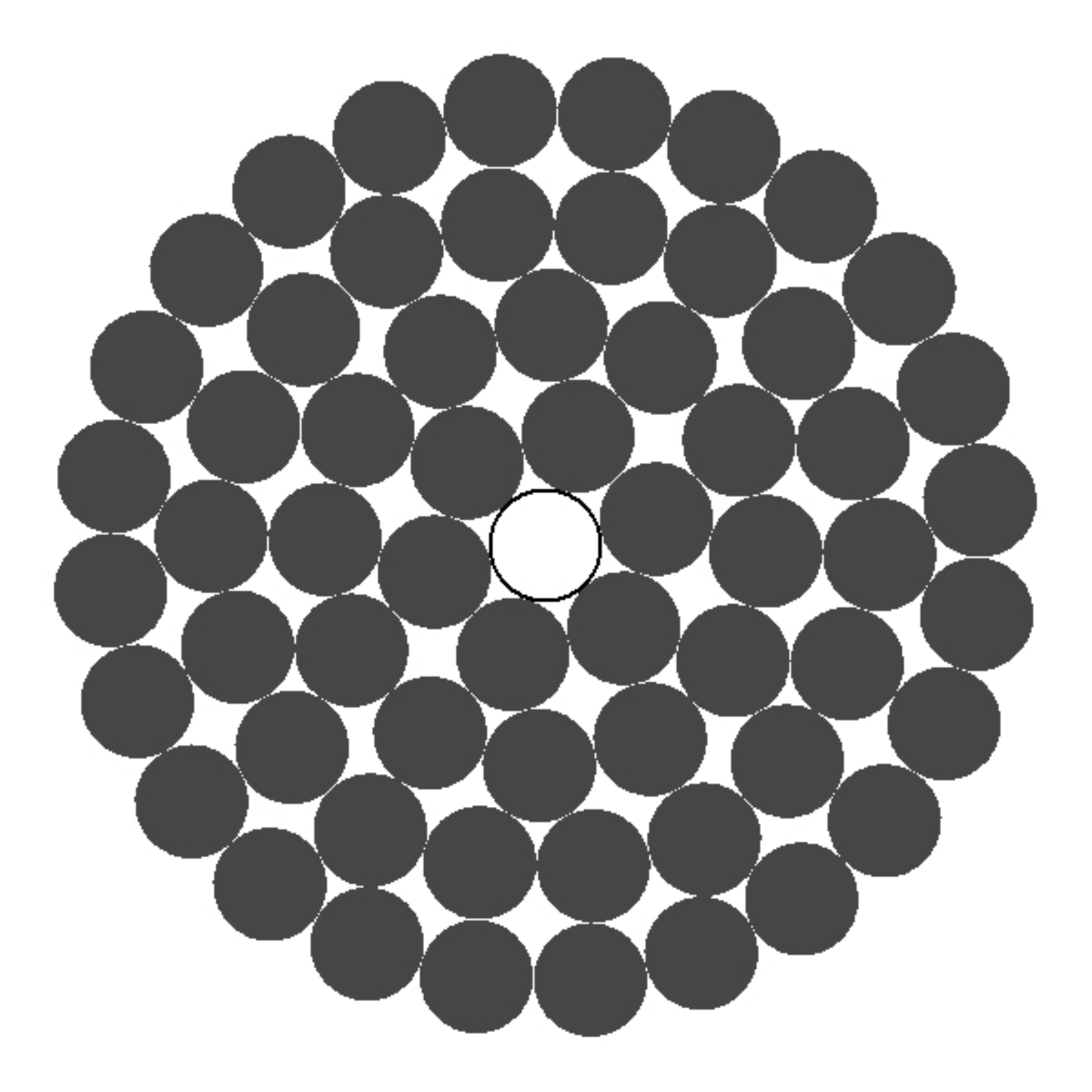}} \\
\subfigure[]{\includegraphics[width=2.0in]{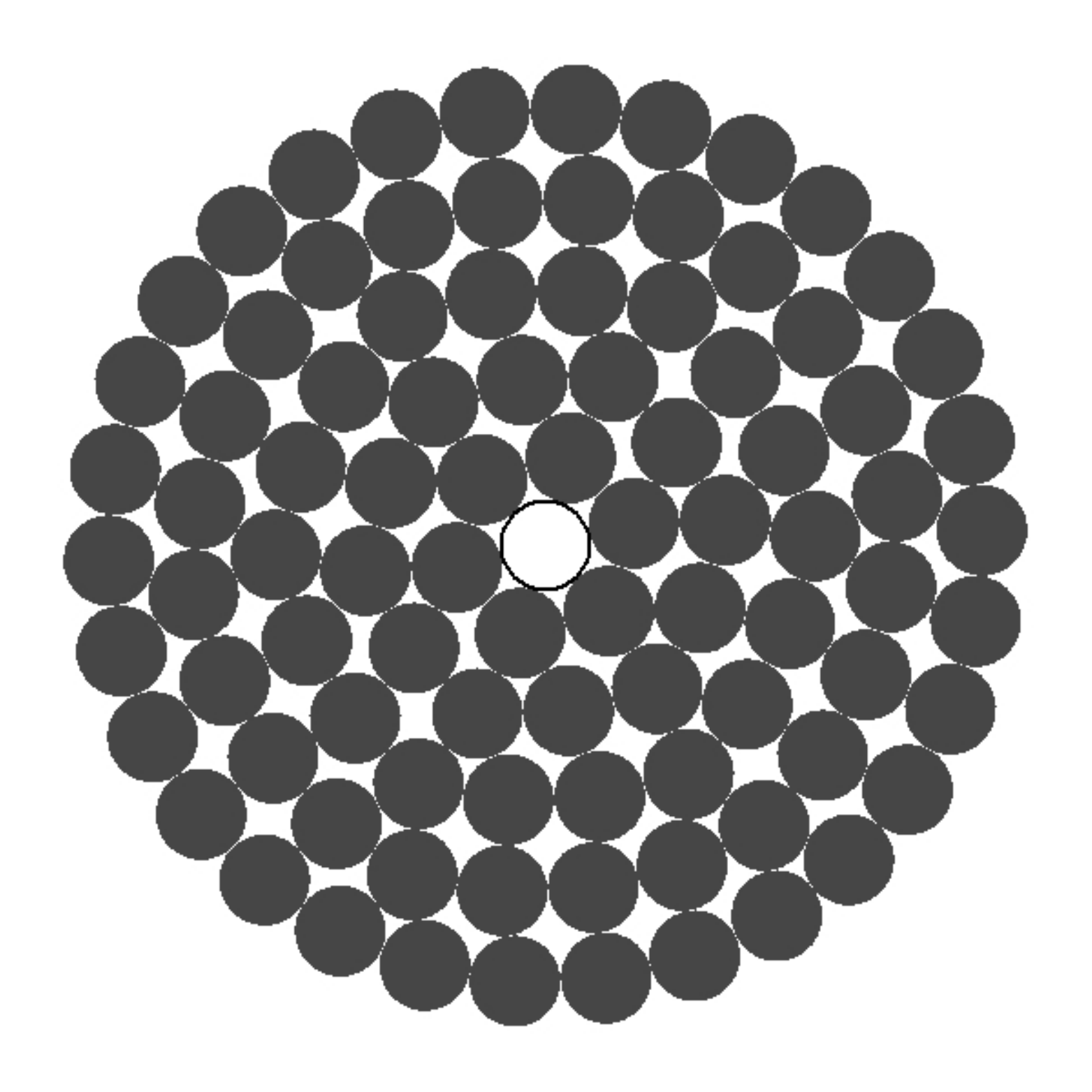}} \\
\subfigure[]{\includegraphics[width=2.0in]{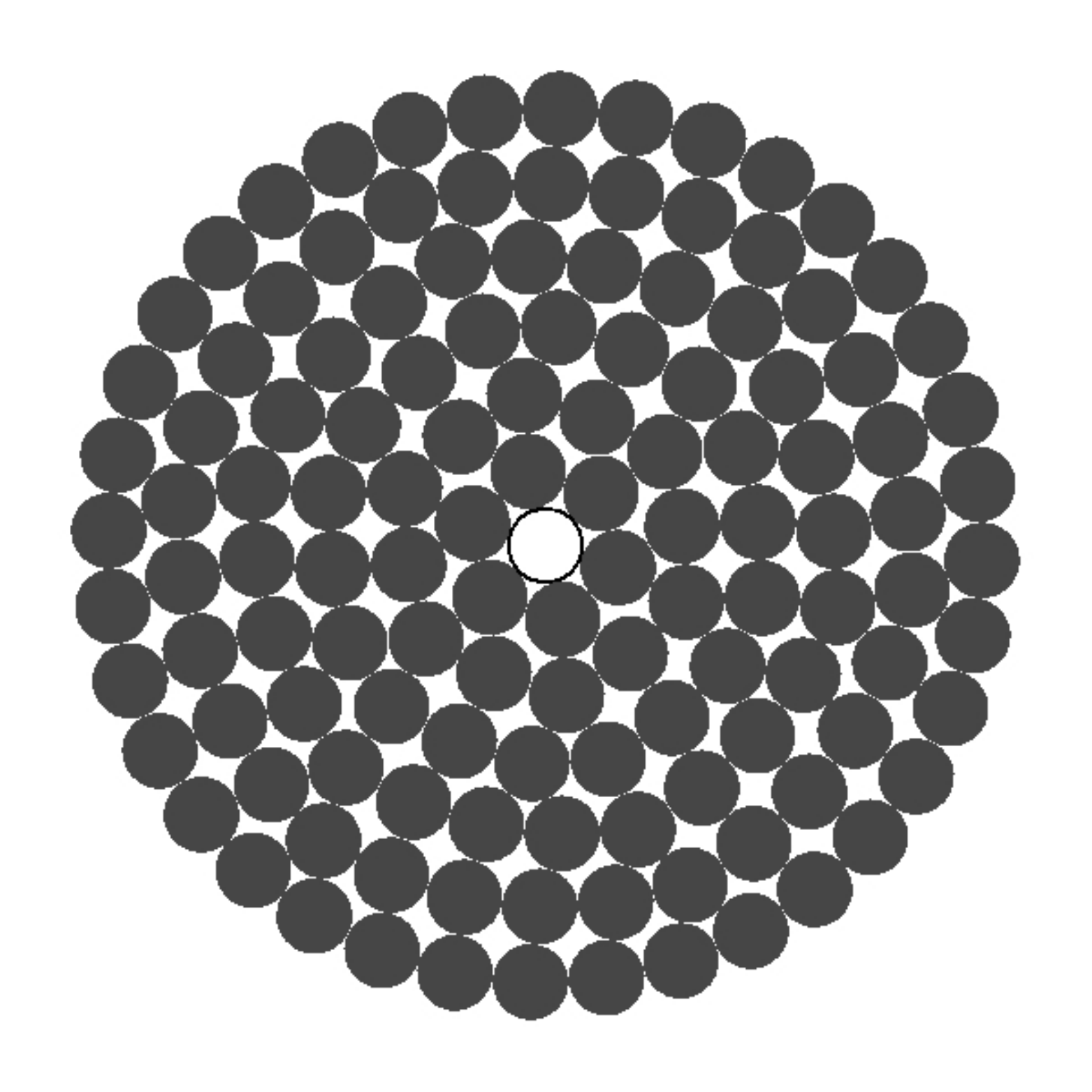}}
\caption{Curved hexagonal packings that are also DLP optimal packings. (a) $N=60$, $R_{min}(60) = 3.830649\dots$, point group $C_{6h}$. (b) $N=90$, $R_{min}(90) = 4.783386\dots$, point group $C_{6h}$. (c) $N=126$, $R_{min}(126) = 5.736857\dots$, point group $C_{6h}$.}
\label{curvedHex}
\end{figure}

For $N > 10$, we know of no packings that have been proved optimal. However, for certain $N$, previous studies have found packings that we conjecture to be DLP optimal packings. For $N+1$ equal to hexagonal number $3k(k+1)+1$, $k \geq 1$ an integer, Lubachevsky and Graham \cite{LG1997a} found a class of packings called \textit{curved hexagonal} packings that they conjectured to be the densest packings of $N+1$ identical nonoverlapping disks within an encompassing disk for $k=1\dots5$. The characteristics of this class include that each packing has a disk fixed at the center of the encompassing disk, and that all curved hexagonal packings belong to point group $C_{6h}$, meaning that they are invariant under $60^{\circ}$ rotation or inversion through the origin. Further, for $k \geq 4$, for each $k$ there are a finite number of degenerate packings of equal density (for $1 \leq k \leq 3$, there is a unique densest packing). The degenerate packings are chiral; each ring of disks, or disks that share a common radius, beginning with the fourth ring (the fourth farthest from the center), can be angularly oriented in more than one distinct fashion relative to the preceding ring. By reorienting rings, the degenerate packings for a given $k$ can be generated from one another. Figure \ref{curvedHex} depicts curved hexagonal packings for $N=60$, $90$ and $126$.

DLP optimal packings for $N$ disks are equivalent to the densest packings of $N+1$ disks enclosed in an encompassing disk when one of the disks is fixed at the center. For $N=6$, $18$, $36$, $60$ and $90$ ($k=1\dots5$), we find that the curved hexagonal packings are the only DLP optimal packings, in support of the conjecture of Lubachevsky and Graham that curved hexagonal packings are the densest packings up to $k=5$. Further, for $N=126$ ($k = 6$), we also find that curved hexagonal packings are the only DLP optimal packings, indicating that though there are packings of $127$ unconstrained disks within an encompassing disk denser than the curved hexagonal packings (as were found by Lubachevsky and Graham \cite{LG1997a}), the curved hexagonal packings remain the densest packings of $126$ disks around a fixed central disk. This is not the case for $N=168$ ($k = 7$), as we find DLP optimal packings with higher density, such as the $N=168$ packing to be shown later in the top panel of Fig. \ref{bigDiff}.

\subsection{Wedge hexagonal packings}
\begin{figure}[!ht]
\centering
\subfigure[]{\includegraphics[width=2.0in]{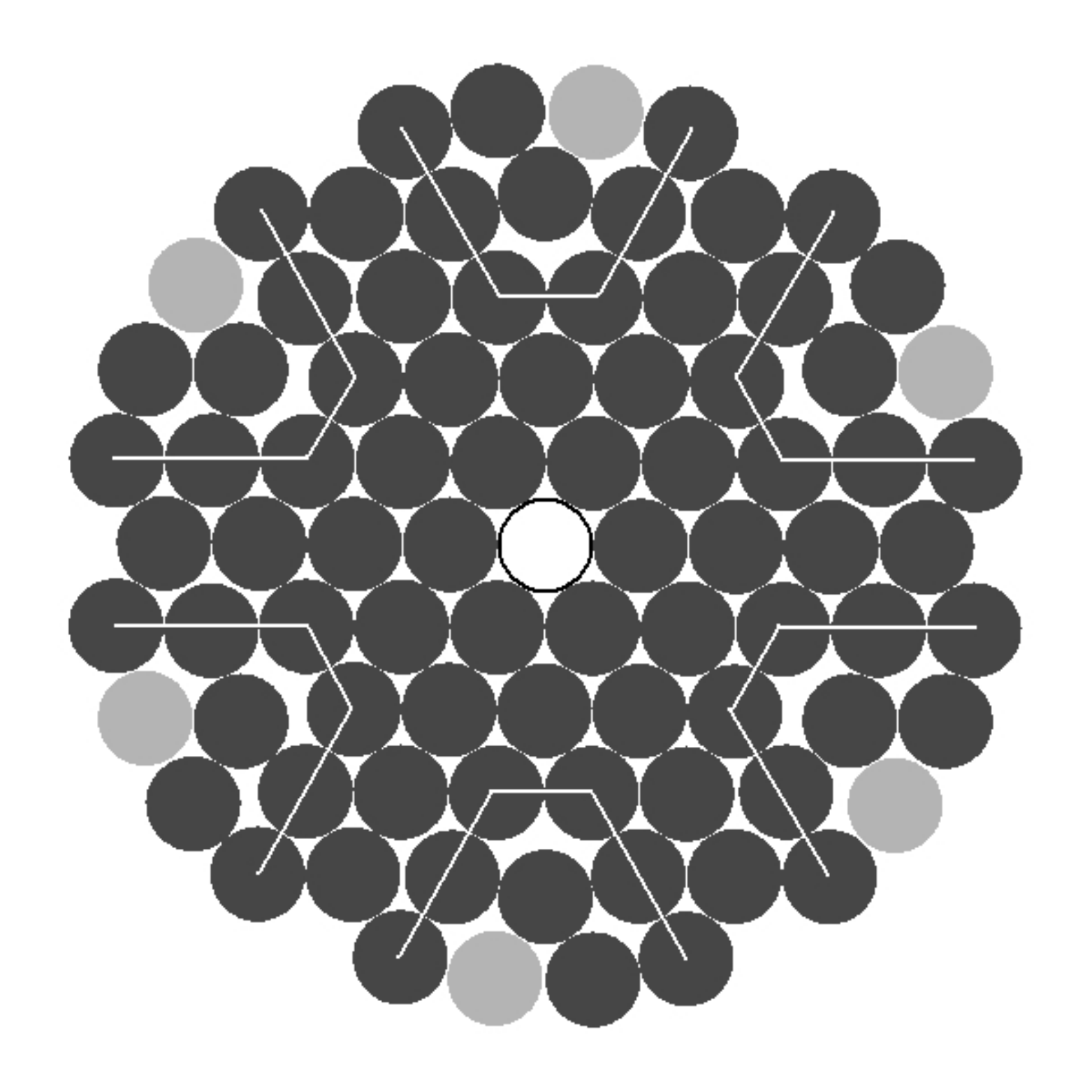}} \\
\subfigure[]{\includegraphics[width=2.0in]{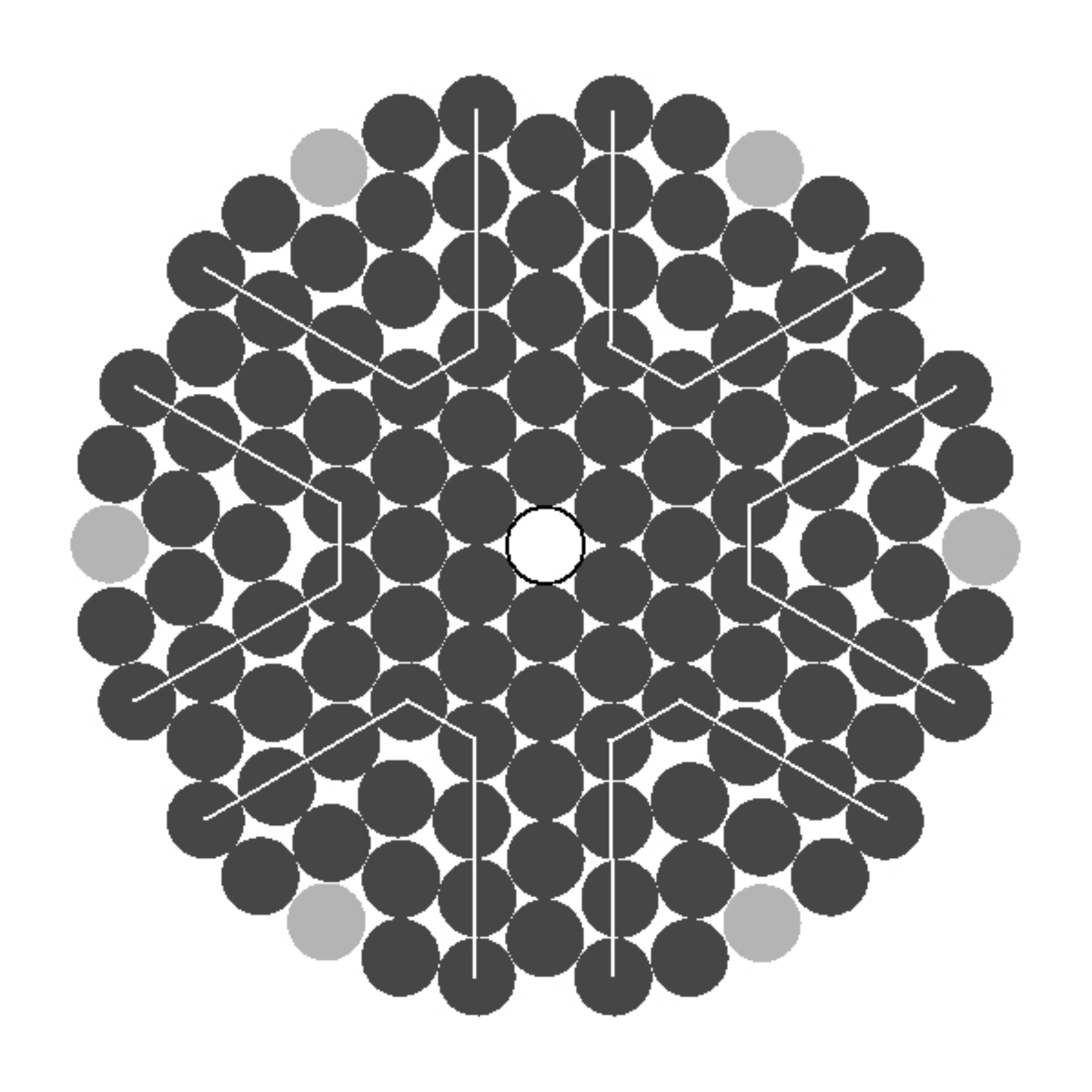}} \\
\subfigure[]{\includegraphics[width=2.0in]{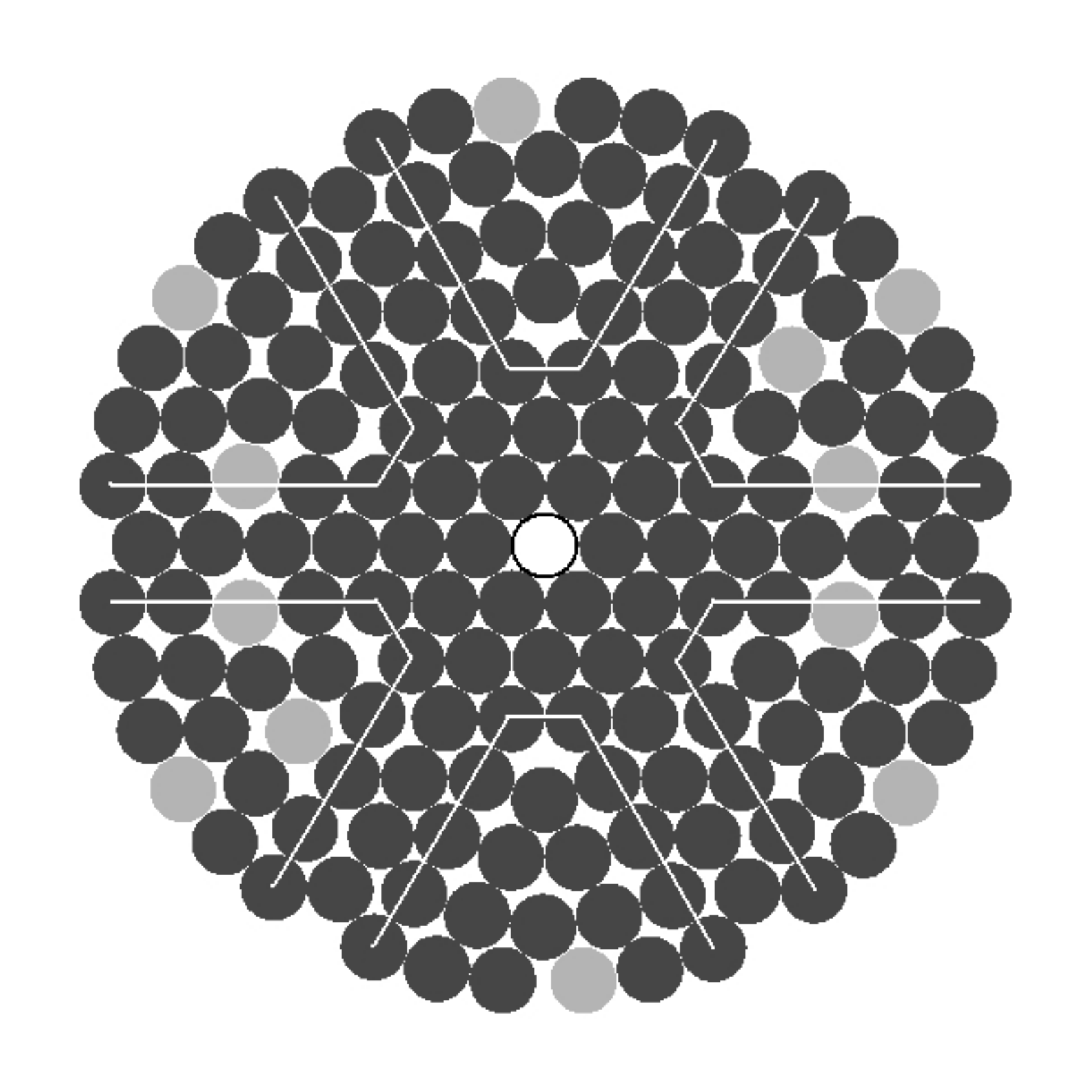}}
\caption{DLP optimal packings that are minor deviations from ``wedge hexagonal'' packings. (a) $N=84$, $R_{min}(84) = 4.581556\dots$, $(k,p,a) = (3,2,1)$, point group $C_i$. (b) $N=120$, $R_{min}(120) = 5.562401\dots$, $(k,p,a) = (3,3,1)$, point group $C_i$. (c) $N=162$, $R_{min}(162) = 6.539939\dots$, $(k,p,a) = (3,4,1)$, point group $C_i$.}
\label{wedgeHexA}
\end{figure}

Another class of packings, previously unidentified, contain a subset of disks with centers arranged on the sites of the triangular lattice and the remainder arranged in six ``wedges''. We hereafter term such packings \textit{wedge hexagonal} packings. Wedge hexagonal packings are not DLP optimal packings when arranged symmetrically (point group $D_{6h}$); however, minor deviations from perfect symmetry in a wedge hexagonal packing can produce a DLP optimal packing. Figure \ref{wedgeHexA} depicts such DLP optimal packings for $N=84$, $120$ and $162$. Lines to guide the eye have been drawn on the three optimal packings in Fig. \ref{wedgeHexA}.

In a wedge hexagonal packing, the subset of disks with centers arranged on the sites of the triangular lattice contains two parts; a regular hexagonal core of hexagonal number $3k(k+1) + 1$ disks, with $k \geq 3$ odd; and six `branches' composed of $(pk-a)$ disks, with $p \geq 2$ and $a \geq 1$ integers, extending from each of the vertices of the core regular hexagon. The branches are $k$ disks wide and $p$ disks long, with $a$ of the farthest disks removed such that the end of the branch approximates a circle (as opposed to the point of a triangle). The six wedges are arranged roughly as $p(p+1)/2$ bowling pins and lie in between the branches, with each of the six `lead pins' placed at the midpoint of each side of the core hexagon. The DLP optimal packings in Fig. \ref{wedgeHexA} are the wedge hexagonal packings, with minor deviations in the positions of some disks, for $(k,p,a) = (3,2\!-\!4,1)$.

\begin{figure}[!ht]
\centering
\subfigure[]{\includegraphics[width=2.0in]{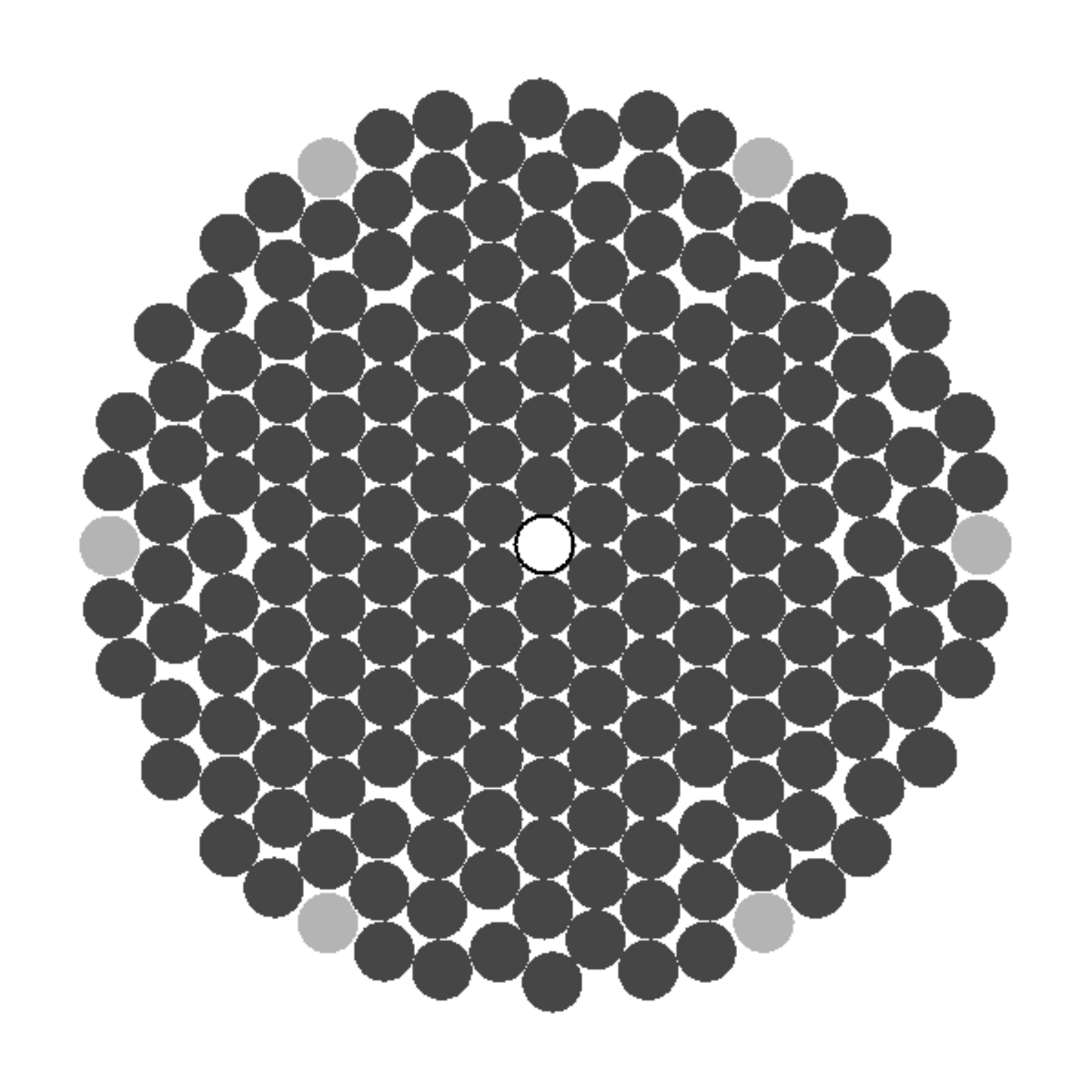}} \\
\subfigure[]{\includegraphics[width=2.0in]{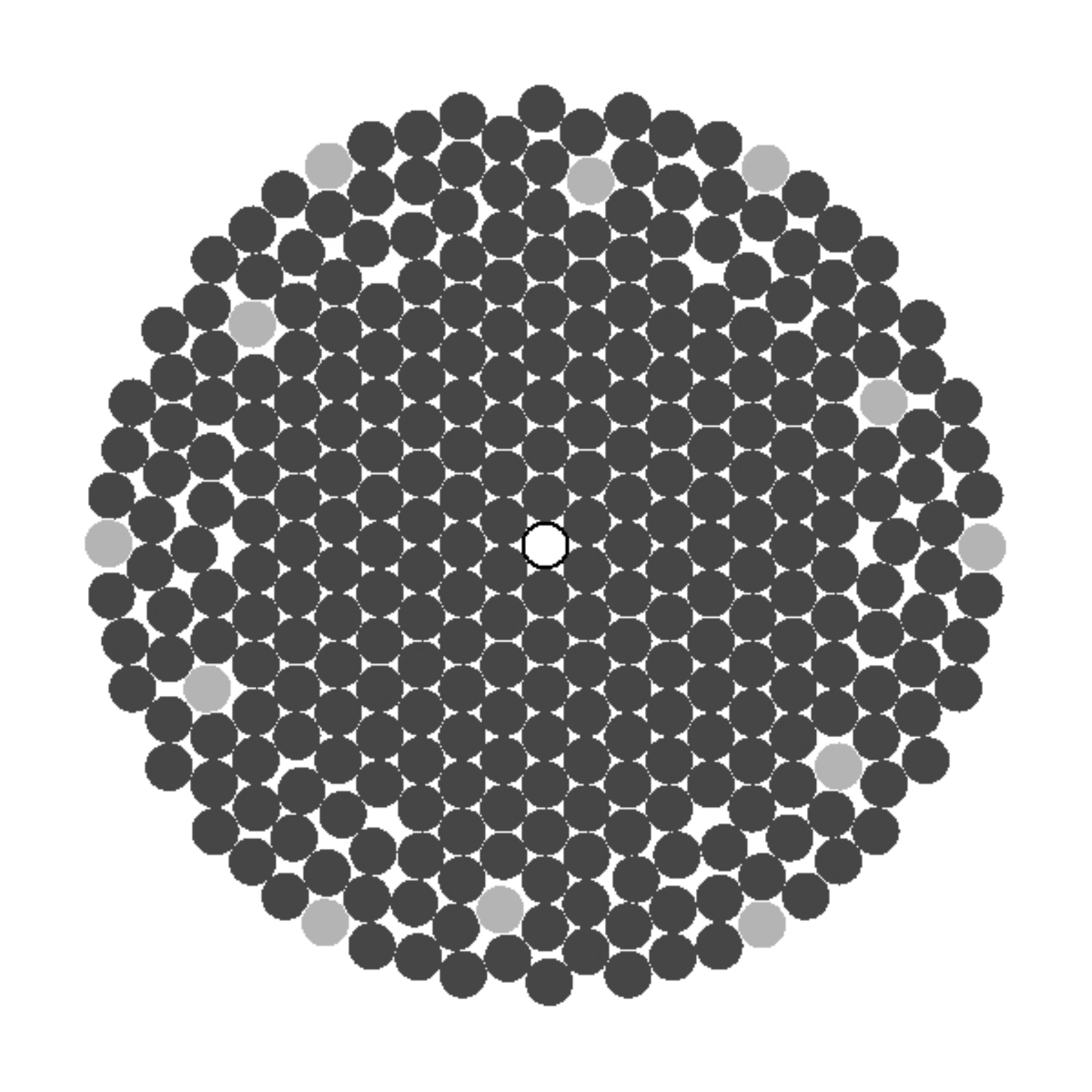}}
\caption{DLP optimal packings that are minor deviations from the ``wedge hexagonal'' packings. (a) $N=198$, $d=2$. $R_{min}(198) = 7.201130\dots$, $(k,p,a) = (5,3,3)$, point group $C_{i}$. (b) $N=312$, $R_{min}(312) = 9.141107\dots$, $(k,p,a) = (7,3,3)$, point group $C_i$.}
\label{wedgeHexB}
\end{figure}

In general, the deviations necessary to produce a DLP optimal packings from a wedge hexagonal packing occur in the branches, and to a lesser degree, the wedges of the packing, while the core regular hexagon retains perfect six-fold symmetry. The deviations required differ for each wedge hexagonal packing, but from our observations produce packings where the backbone maintains inversion symmetry through the origin, where the backbone of a packing is defined as the packing excluding the rattlers. Such deviations can be seen in the branches and wedges in the DLP optimal packings for $N=198$ and $N=312$ in Fig. \ref{wedgeHexB}, which correspond to the (slightly altered) wedge hexagonal packings with $(k,p,a) = (5,3,3)$ and $(7,3,3)$, respectively.

\subsection{DLP optimal packings with high symmetry}
Many DLP optimal packings exhibit symmetries other than inversion symmetry through the origin as exhibited by the altered wedge hexagonal packings shown in Figs. \ref{wedgeHexA} and \ref{wedgeHexB}. These symmetries include perfect bond orientational order, invariance under rotation through an angle, and invariance under reflection across an axis. A list of packing point group, alongside other packing properties such as $R_{min}(N)$ value, of all of the DLP optimal packings depicted in this work appears in Appendix \ref{tables}.

\begin{figure}[!ht]
\centering
\subfigure[]{\includegraphics[width=2.0in, viewport = 30 95 595 680,clip]{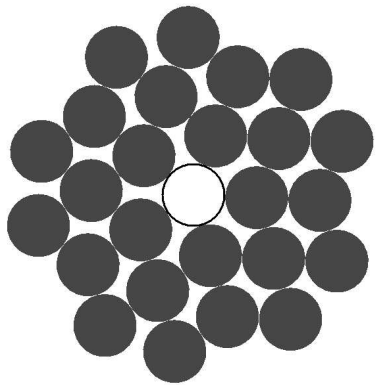}} \\
\subfigure[]{\includegraphics[width=2.0in, viewport = 30 95 595 680,clip]{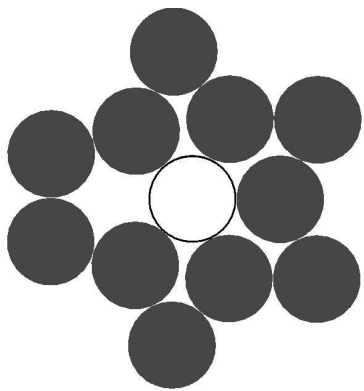}} \\
\subfigure[]{\includegraphics[width=2.0in, viewport = 30 95 595 680,clip]{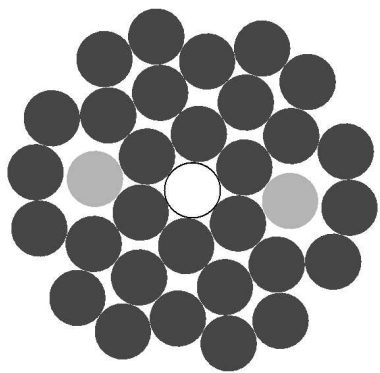}}
\caption{Three examples of DLP optimal packings incorporating interesting symmetry elements. (a) $N=25$, $R_{min}(25) = 2.497212\dots$, point group $D_{5h}$. (b) $N=11$, $R_{min}(11) = 1.685854\dots$, point group $C_{2v}$. (c) $N=32$, $R_{min}(32) = 2.794164\dots$, point group $D_{2h}$.}
\label{oddSyms}
\end{figure}

Perfect five-fold symmetry, disallowed to regular infinite crystals, is exhibited by three of the optimal packings studied. Five-fold rotational symmetry is present in the $N=15$ (Fig. \ref{DLP15pic}), $N=10$ (bottom panel of Fig. \ref{10opt}) and $N=25$ (top panel of Fig. \ref{oddSyms}) packings. The $N=25$ optimal packing also has perfect five-fold bond orientational order, evident in that all nearest-neighbor disk pairs are at one of five angles relative to any fixed coordinate system. Additionally, it is of note that the $N=25$ packing may be tiled by $15$ identical rhombuses of acute angle $72^{\circ}$ with vertices placed at disk centers, where the rhombus of acute angle $72^{\circ}$ is known to be the `thicker' of the two types of rhombus present in a Penrose tiling \cite{Penrose1974a}. The top panel of Fig. \ref{jmolPics}, a diagram of the contact network for the $N=25$ optimal packing, depicts these rhombuses.

Two other optimal packings incorporating interesting symmetry elements are the $N=11$ and $N=32$ packings (center and bottom panels of Fig. \ref{oddSyms}, respectively). The packing depicted for $N=11$ belongs to symmetry group $C_{2v}$ and appears to be the unique DLP optimal packing of $11$ disks. The packing for $N=32$ is one of an infinity of possible packings due to the presence of two rattlers; however, the backbone of the packing has reflection symmetry across two axes and inversion symmetry through the origin and belongs to symmetry group $D_{2h}$.

\begin{figure}[!ht]
\centering
\includegraphics[width=2.0in]{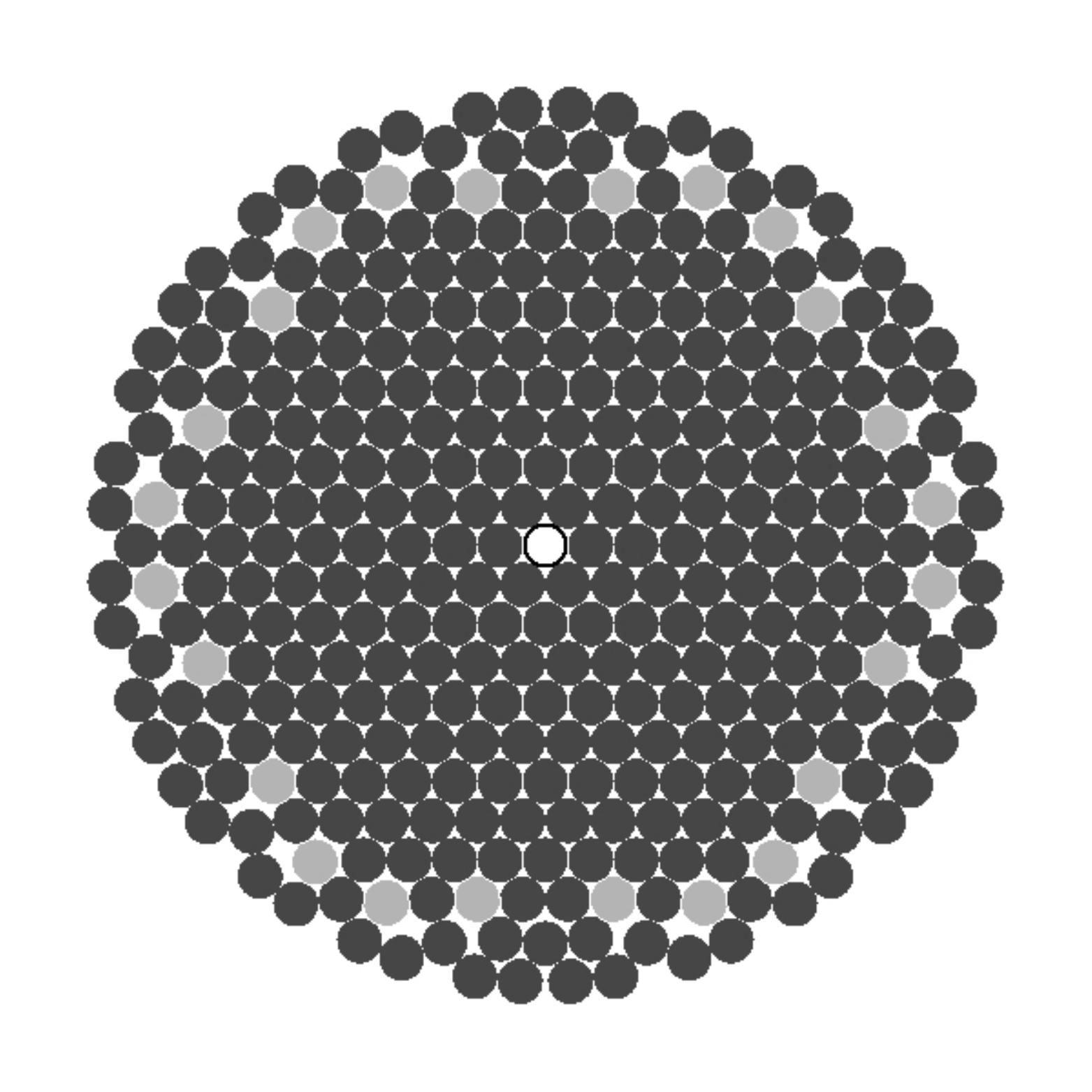}
\caption{DLP optimal packing for $N=348$, $R_{min}(348) = 9.620709\dots$, belonging to point group $D_{6h}$.}
\label{manySpheres}
\end{figure}

Though five-fold symmetry may be limited to packings with small $N$, high symmetry in general is not. For example, the backbone of the optimal packing with the largest $N$ presented here, $N=348$, has six-fold rotation symmetry and belongs to point group $D_{6h}$. Figure \ref{manySpheres} depicts the $N=348$ optimal packing, which contains $24$ rattlers. 

\subsection{Unusual features in select optimal packings}

\begin{figure}[!ht]
\centering
\subfigure[]{\includegraphics[width=2.0in, viewport = 30 95 595 680,clip]{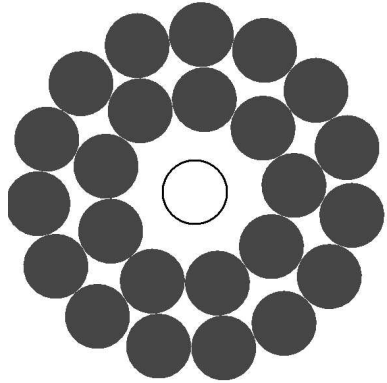}} \\
\subfigure[]{\includegraphics[width=2.0in, viewport = 30 95 595 680,clip]{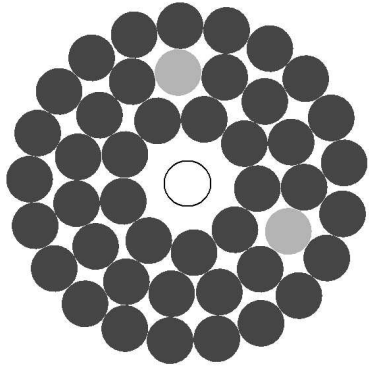}} \\
\subfigure[]{\includegraphics[width=2.0in, viewport = 30 95 595 680,clip]{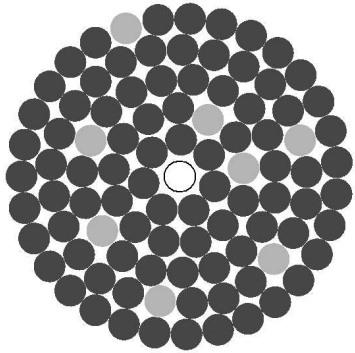}}
\caption{Three examples of DLP optimal packings with circular cavities about the fixed central disk. (a) $N=24$, $R_{min}(24) = 2.425256\dots$, point group $D_{3h}$. (b) $N=45$, $R_{min}(45) = 3.374023$, point group $C_1$. (c) $N=95$, $R_{min}(95) = 4.958096$, point group $C_1$.}
\label{circCavities}
\end{figure}

A prevalent feature found in many of the DLP optimal packings studied is a cavity consisting of a ring of disks enclosing, but not contacting, the fixed central disk. The counterintuitive presence of this feature is related to the aforementioned fact that in any dimension $d$, no spherical window of radius $R \leq \tau$ centered on a central nonoverlapping sphere of unit diameter may encircle more sphere centers than the number (plus one) that can be placed on the encircling sphere's surface. Circular cavities around the fixed central disk appear in many DLP optimal packings, including those for $N=24$, $45$ and $95$ disks, as illustrated in Fig. \ref{circCavities}. There are $9$, $8$ and $7$ disks, respectively, forming the walls of the cavities in the three packings in Fig. \ref{circCavities}.

\begin{figure}[ht]
\centering
\subfigure[]{\includegraphics[width=2.0in]{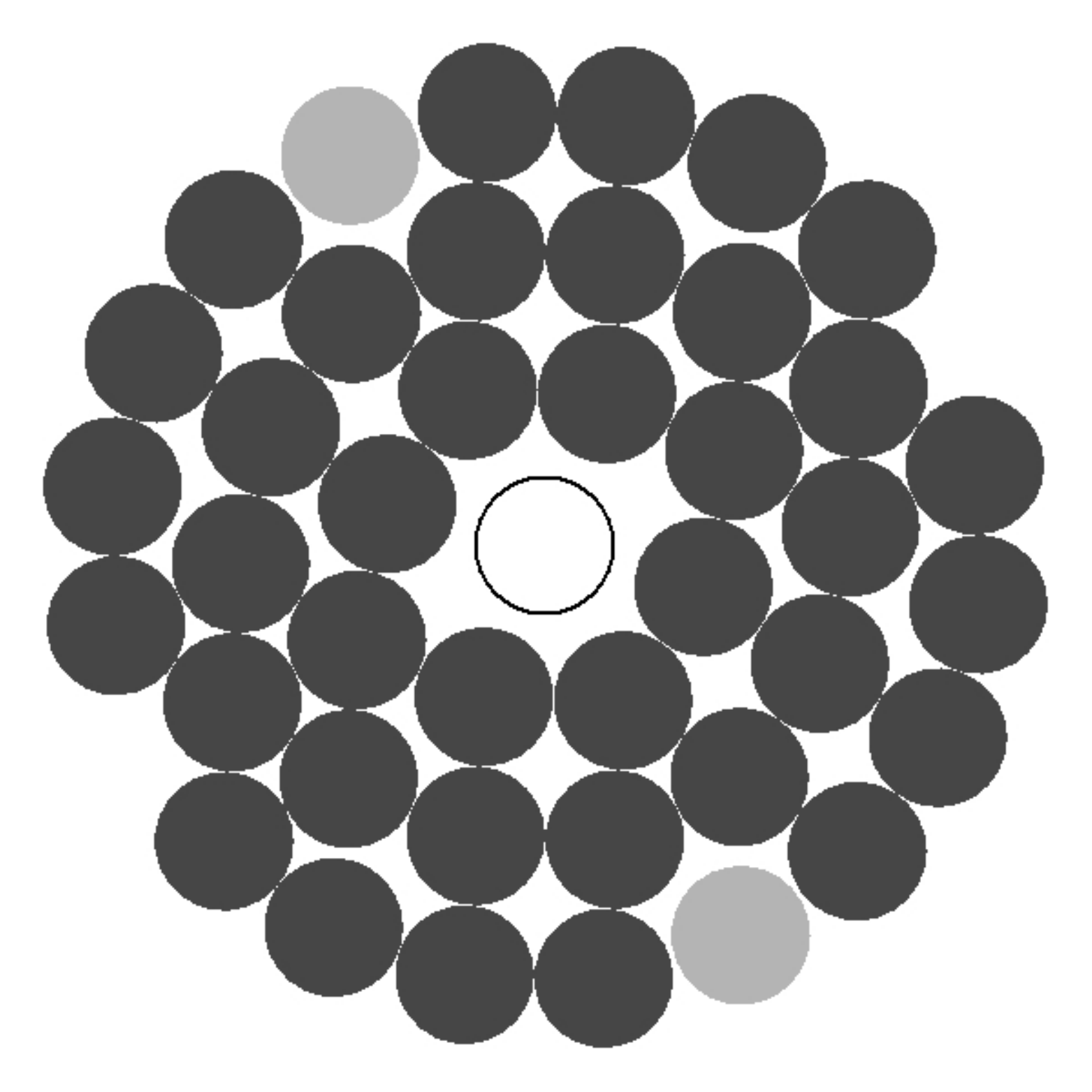}} \\
\subfigure[]{\includegraphics[width=2.0in]{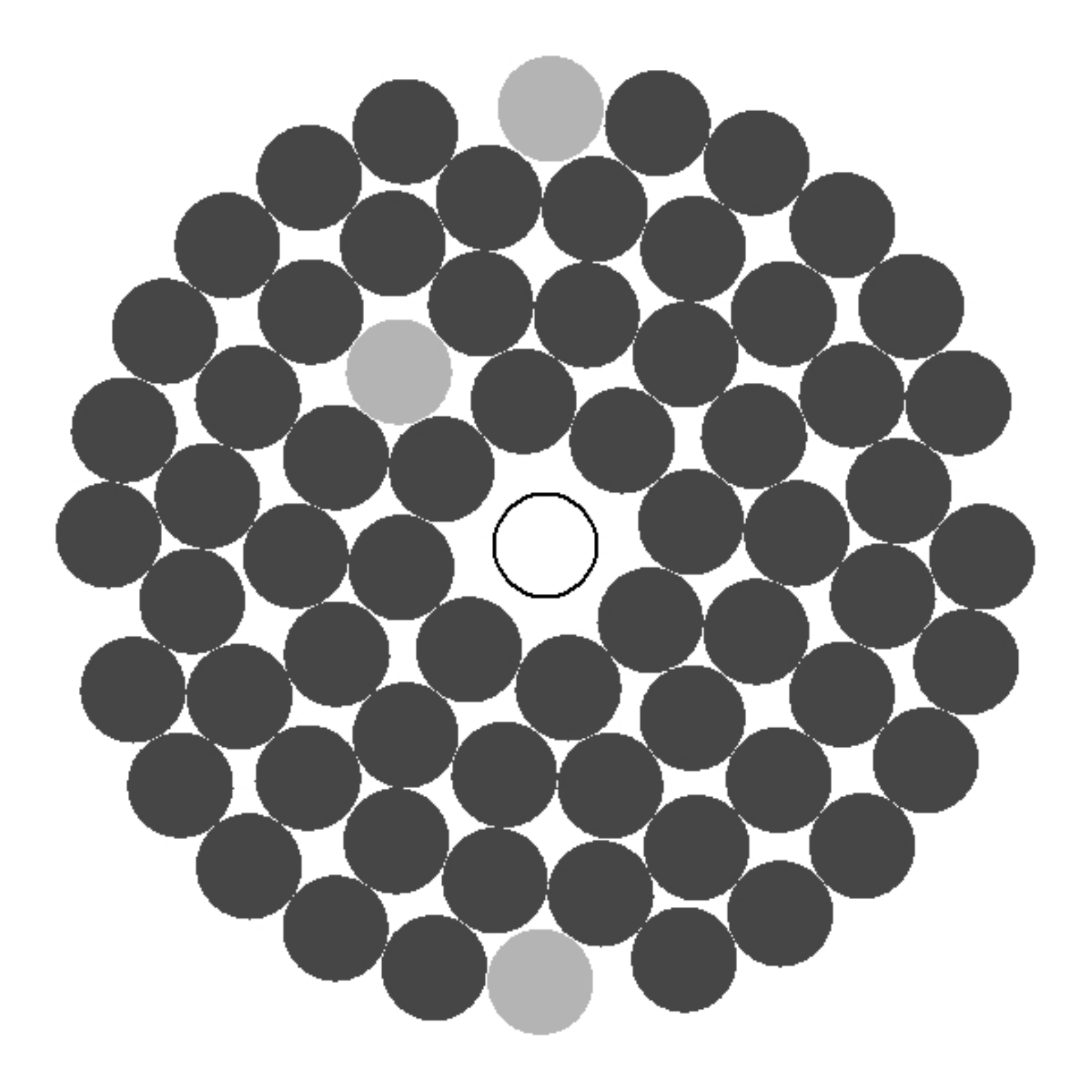}}
\caption{DLP optimal packings consisting of layers of distorted rings, or `eyes', enclosing the central disk. (a) $N=40$, $R_{min}(40) = 3.136712\dots$, point group $D_{2h}$. (b) $N=66$, $R_{min}(66) = 4.104997$, point group $C_1$.}
\label{eyePackings}
\end{figure}

\begin{figure}[!ht]
\centering
\subfigure[]{\includegraphics[width=2.0in, viewport = 30 125 595 635,clip]{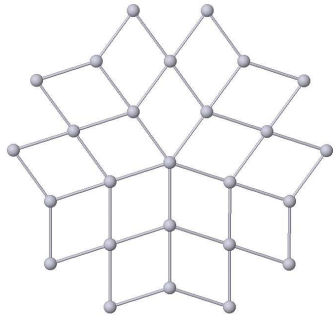}} \\
\subfigure[]{\includegraphics[width=2.0in, viewport = 30 125 595 635,clip]{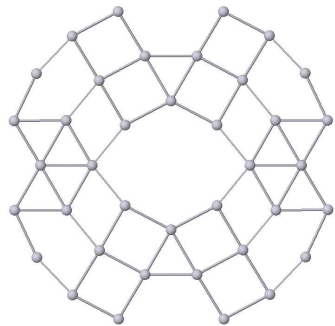}}
\caption{Diagrams \cite{jmolRef} of contact networks for the $N=25$ (a) and $N=40$ (b) optimal packings, with point group symmetries $D_{5h}$ and $D_{2h}$, respectively.}
\label{jmolPics}
\end{figure}

Two particularly notable DLP optimal packings that include a cavity enclosing the fixed central disk are the $N=40$ and $N=66$ packings shown in Fig. \ref{eyePackings}.
These packings are composed of layers of distorted rings, where the distorted rings appear as eye-like closed curves of varying curvature with each successive layer from the center more circular than the last. It is curious that even though the only shape imposed upon the packings, in the form of the encompassing disk and the disks themselves, is circular, that an optimal packing incorporating distorted rings emerges for these numbers of disks. The bottom panel of Fig. \ref{jmolPics} is a diagram of the contact network for the backbone of the $N=40$ optimal packing.

The presence of these cavities about the central disk leads to an interesting counterintuitive result. Suppose that in a binary liquid of nonoverlapping disks of unit diameter, one species of disk is endowed with an attractive square well potential. The potential of the ``attractive'' disks, present in the dilute limit in comparison to the ``nonattractive'' disks, acts on the centers of the ``nonattractive'' disks extending only to a distance just larger than any of the $R_{min}(N)$ presented in Figs. \ref{circCavities} or \ref{eyePackings}. If the depth (strength) of this square well were made arbitrarily large, then the result would seem paradoxical: unbounded attraction to the disks, in a minimal energy configuration, would eliminate contact with these disks.

This effect in such a binary liquid of disks requires that the pair correlation function depicting the probability density of finding the centers of a given number of ``nonattractive'' disks a certain distance from the centers of ``attractive'' disks be zero from $r=0$ to $r = R_0 > 1$ a specified distance in excess of the diameter of the disks. None of the currently available pair correlation function theories are able to predict this effect, including the crucial dependence of $R_0$ on $R_{min}(N)$, because the underlying approximations of the currently available theories cannot account for the basic many-body geometrical features involved. 

\subsection{Imperfect symmetry}
\begin{figure}[!ht]
\centering
\subfigure[]{\includegraphics[width=2.0in, viewport = 30 95 595 680,clip]{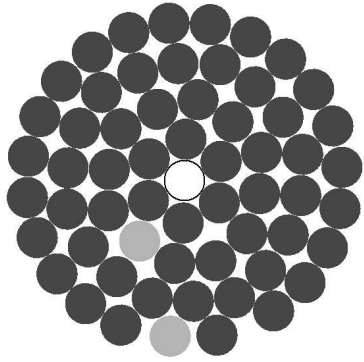}} \\
\subfigure[]{\includegraphics[width=2.0in, viewport = 30 95 595 680,clip]{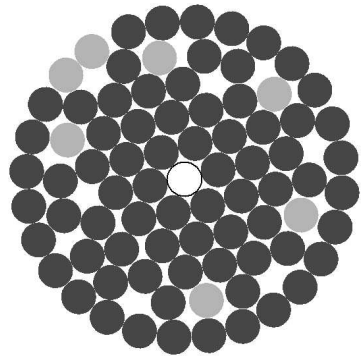}} \\
\subfigure[]{\includegraphics[width=2.0in, viewport = 30 95 595 680,clip]{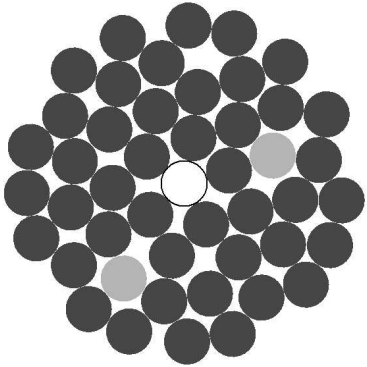}}
\caption{DLP optimal packings that exhibit imperfect symmetry. (a) $N=59$, $R_{min}(59) = 4.824374$, point group $C_1$. (b) $N=80$, $R_{min}(80) = 4.514170$, point group $C_1$. (c) $N=46$, $R_{min}(46) = 3.414304$, point group $C_1$.}
\label{noSym}
\end{figure}

Not all DLP optimal packings exhibit perfect symmetry; for many $N$, a subset of disks in an optimal configuration appear to mimic a symmetric packing, but the packing as a whole exhibits only imperfect symmetry. One situation in which this occurs frequently is when the number of disks $N$ in the optimal packing is close to a different number for which the optimal packing is relatively unusually dense. For example, the optimal $N=59$ packing shown in the top panel of of Fig. \ref{noSym} lacks any of the symmetry elements described above but nonetheless closely resembles the particularly dense $N=60$ curved hexagonal packing (top panel of Fig. \ref{curvedHex}).

Other packings in which imperfect symmetry is present include the $N=80$ packing shown in the center panel of Fig. \ref{noSym}, the disks closer to the center of which are ordered with centers on the sites of a slightly distorted triangular lattice, and the $N=46$ packing shown in the bottom panel of Fig. \ref{noSym}, which has imperfect five-fold symmetry. The $N=46$ packing, along with the $N=45$ packing (center panel of Fig. \ref{circCavities}), together illustrate another finding: that the structure of optimal packings even for consecutive numbers of disks can vary substantially.

\subsection{Surface effects}
\begin{figure}[!ht]
\centering
\subfigure[]{\includegraphics[width=2.0in]{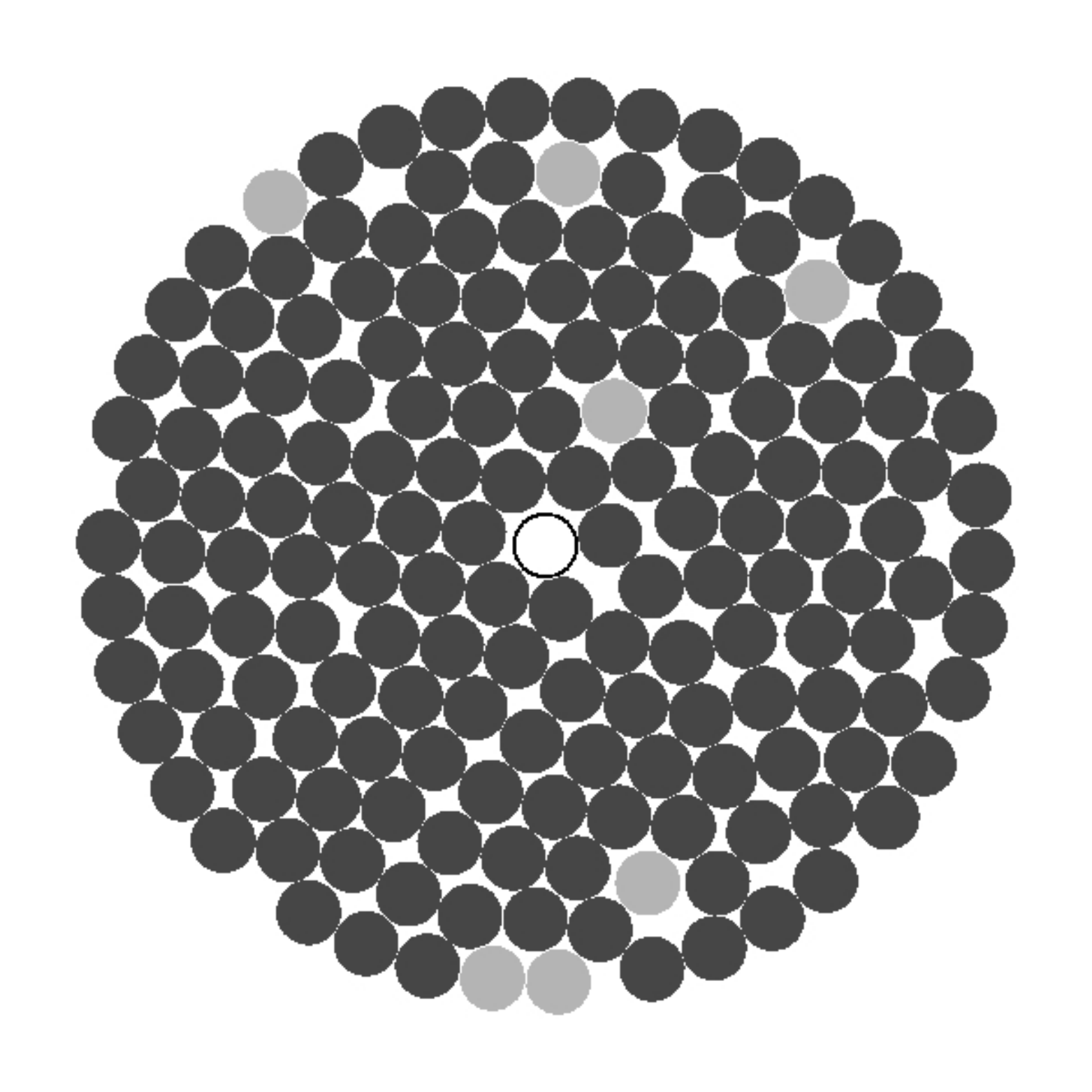}} \\
\subfigure[]{\includegraphics[width=2.0in]{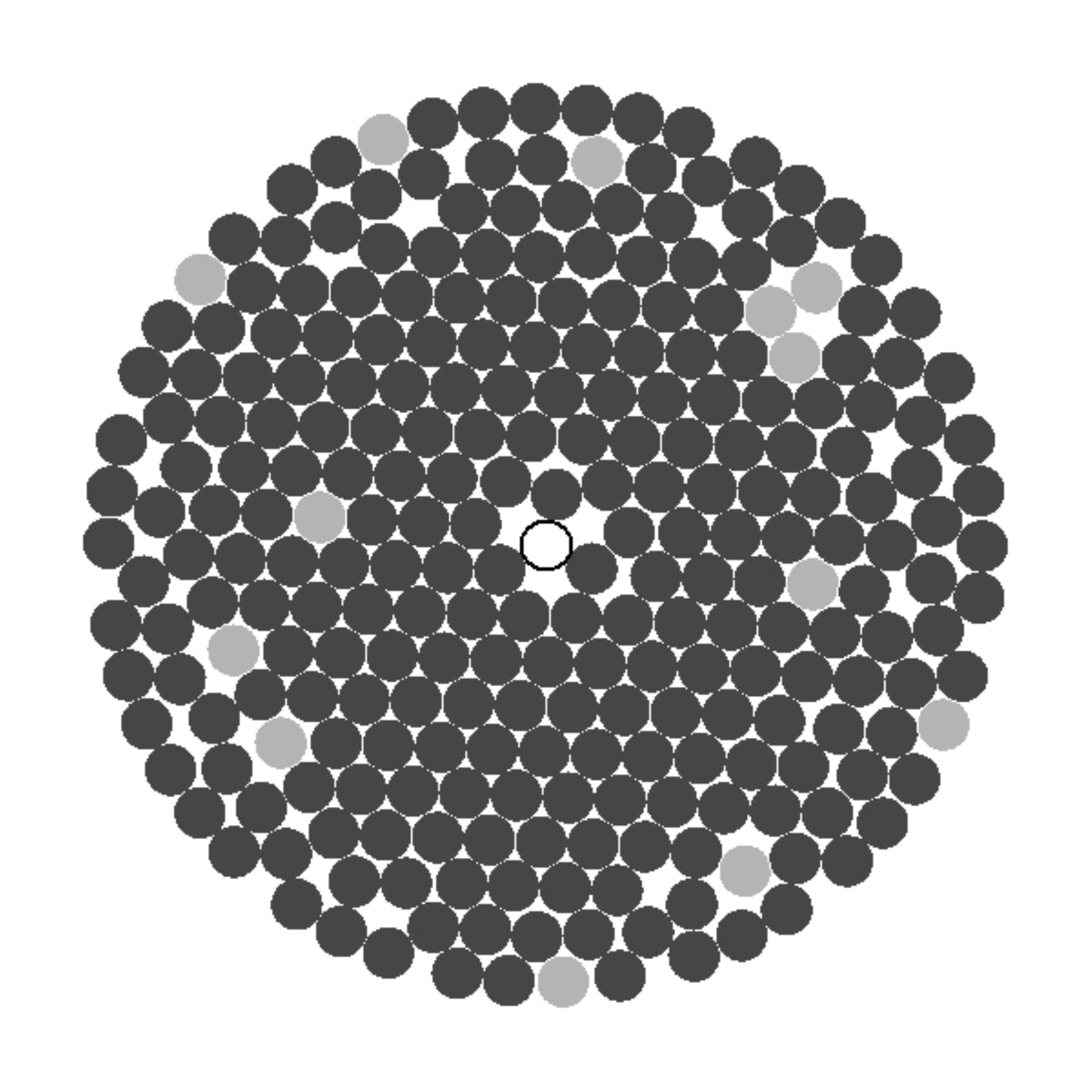}} \\
\subfigure[]{\includegraphics[width=2.0in]{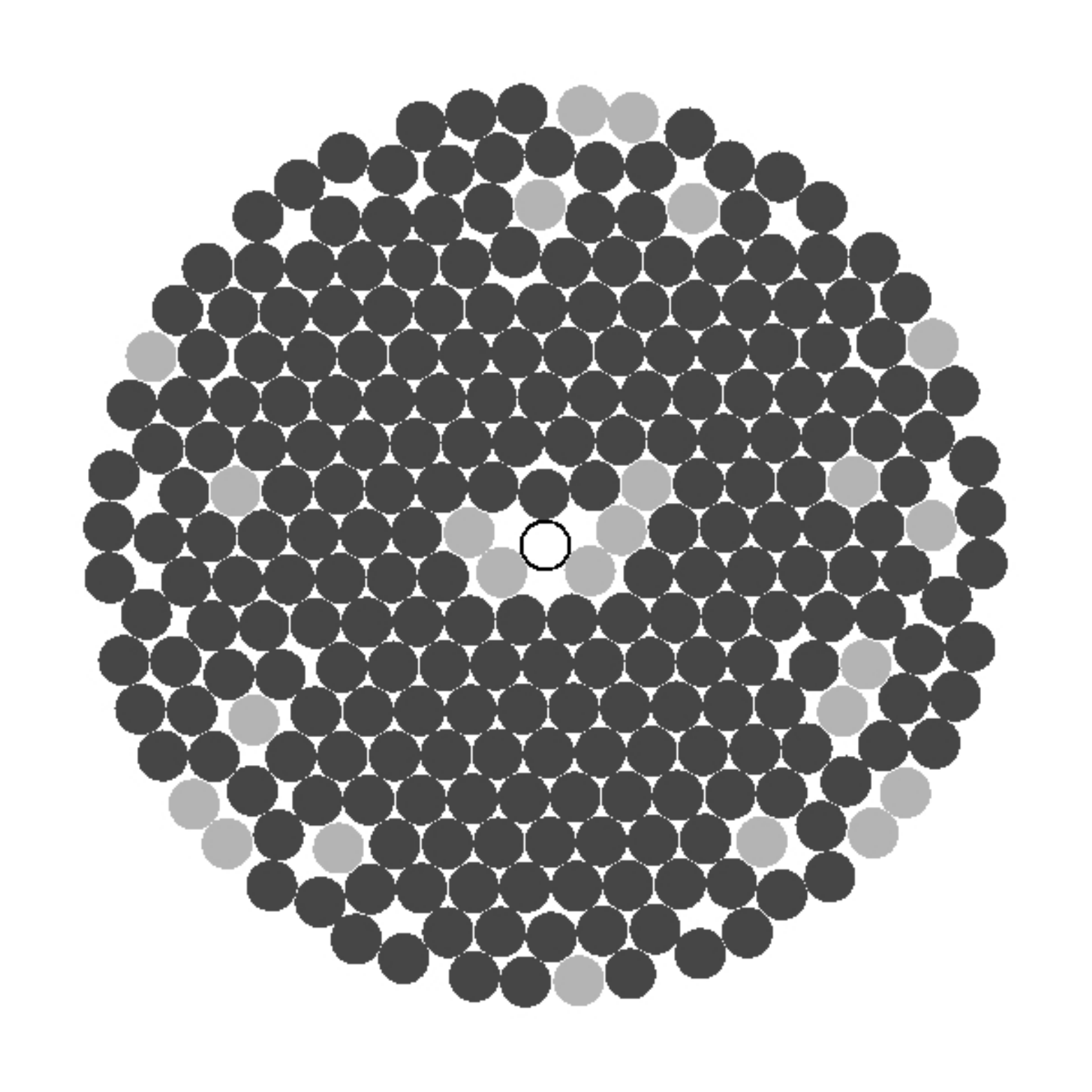}}
\caption{Three packings for which $R_{min}(N)$ is significantly less than the radius of a disk enclosing the centers of $N+1$ disks arranged on the sites of the triangular lattice. (a) $N=168$, $R_{min}(168) = 6.680013$, point group $C_1$. (b) $N=264$, $R_{min}(264) = 8.417769$, point group $C_1$. (c) $N=270$, $R_{min}(270) = 8.497744$, point group $C_1$.}
\label{bigDiff}
\end{figure}

DLP optimal packings with $N$ in the higher range of the packings studied appear, as $N$ increases, to more and more resemble the triangular lattice in the bulk of the packing. Nonetheless, the surface of the packing always deviates significantly from the bulk crystal. In general, the optimal packings at higher $N$ consist of a ``bulk zone'' with disk centers arranged in the triangular lattice surrounded by a ``surface zone'' with disk centers arranged in circular rings. This effect can be seen in all of the optimal packings with $N$ in the higher range of $N$ studied, including in those shown in Figs. \ref{wedgeHexA}, \ref{wedgeHexB}, \ref{manySpheres}, in the center panel of Fig. \ref{noSym} and in Fig. \ref{bigDiff}.

Qualitatively, it appears that the radial width of the surface zone tends to increase with number of disks, though not as fast as the radial width of the bulk. Due to computational time constraints, we were unfortunately not able to quantitatively verify this result at much larger $N$; however, should the observed trend continue, the width of the surface zone would continue to grow as the bulk grows, eventually becoming infinitely large as $N\rightarrow\infty$. This does not imply that at very large $N$ the surface zone would represent a substantial fraction of the total packing; as $N\rightarrow\infty$, the ratio of space covered by the surface zone to the space covered by the bulk zone will still be zero.

For certain optimal packings exhibiting this surface effect, we also find that the optimal radius $R_{min}(N)$ is significantly smaller than the radius of the smallest disk centered on a fixed central disk that can enclose (an additional) $N$ disks with centers arranged on the sites of the triangular lattice. Figure \ref{bigDiff} depicts three optimal packings for which this difference is relatively large, at $N=168$, $264$ and $270$, where the $R_{min}(N)$ differ by $0.248190$, $0.128616$ and $0.162510$ from the respective radii for the triangular lattice configuration, $7$, $\sqrt{73}$ and $\sqrt{75}$. Each of the packings in Fig. \ref{bigDiff} also displays its own interesting features, including a close resemblance on the left of the image of the $N=168$ packing to the $(3,4,1)$ wedge hexagonal packing, and imperfect three-fold rotational symmetry for the $N=264$ and $N=270$ packings. 

\section{Conclusions and discussion}
The DLP problem is a local packing problem that in certain limits encompasses both the infinite sphere packing and kissing number problems. DLP optimal packings exhibit a wide variety of symmetries, vary significantly from packings with disk centers placed on the sites of a triangular lattice, and often vary significantly with consecutive $N$. Two local packing classes, the curved hexagonal and wedge hexagonal packings, lead to the densest or very dense packings of disks over the range of $N$ studied.

The optimal radii $R_{min}(N)$ corresponding to a packing of $N$ spheres in ${\mathbb R}^d$ form a realizability condition on functions that are candidates to be the pair correlation function $g_2(r)$ of a statistically homogeneous and isotropic packing of spheres. This realizability condition incorporates more information than is included in the structure factor and pair correlation function nonnegativity conditions alone. Though the condition discussed here only applies to packings of identical spheres, equivalent $Z_{max}(R)$ functions for packings of differentiated non-spherical objects can be found and corresponding realizability conditions imposed.

The function $Z_{max}(R)$ can also be employed in the two ways discussed to bound from above the packing fraction of an infinite sphere packing in dimension $d$. Similarly, $Z_{max}(R)$ for differentiated non-spherical objects can be employed to form upper bounds on corresponding infinite packing maximal packing fractions.

Our work has direct application to packaging, particularly to problems involving identical nonoverlapping disks within a circular boundary. Further, as is discussed in Appendix \ref{algorithm}, the algorithm we have employed to find the $d=2$ putative DLP optimal packings presented in this work may be modified to study dense packings of $d$-dimensional differentiated objects of various shape within different boundaries. Additionally, our work has implications for nucleation theory and surface physics, particularly in terms of the effects of imposing a circular boundary upon a packing of spheres with centers initially placed on the sites of a triangular lattice. In future work, we expect to investigate these implications and others in more depth.

In a sequel to this paper, we will present and analyze DLP optimal packings and their corresponding $R_{min}(N)$ for three-dimensional spheres over a larger range of $N$. We will catalogue optimal packings of particularly high packing fraction and of unusual symmetry, and we will investigate the possibility of $d=3$ extensions to special $d=2$ classes of packings such as the curved hexagonal and wedge hexagonal packings. We will compare $R_{min}(N)$ values to shell distances in Barlow packings \cite{Barlow1883a}, where the Barlow packings, of which the best-known are face centered cubic (FCC) and hexagonal close packed (HCP) arrangements, all individually achieve the maximal infinite-volume packing fraction for $d=3$, $\phi_*^{\infty} = \pi/\sqrt{18} = 0.740481\dots$.

Packing in ${\mathbb R}^3$ is intrinsically more complicated than in ${\mathbb R}^2$. In ${\mathbb R}^3$, there is a wider range of possibilities for contact coordination, particularly for twelve contacting spheres around a central sphere (with $K_3 = 12$) where there is an infinity of possible configurations, as compared to just one in ${\mathbb R}^2$. Additionally, there is a single optimal infinite-volume packing configuration in ${\mathbb R}^2$, i.e., the triangular lattice, whereas there is an infinite number in ${\mathbb R}^3$, i.e., the Barlow packings.

Preliminary findings indicate that there is less symmetry (as quantified by point groups) present for the majority of DLP optimal packings in ${\mathbb R}^3$ and more variation with $N$. Further, observations suggest that for the same value of $N$ that there are significantly more locally jammed packing configurations in ${\mathbb R}^3$ with radius $R > R_{min}(N)$ than is the case in ${\mathbb R}^2$. This finding has implications for the dynamics of nucleation occuring in pure supersaturated liquids. If a nucleus in a supersaturated liquid of identical nonoverlapping spheres in ${\mathbb R}^3$ is approximated as a group of $N$ densely-packed spheres with centers within distance $R$ of a central sphere, then there are more packing configurations available to a nucleus of radius $R$, with $R$ confined to a small finite range $R > R_{min}(N)$, in ${\mathbb R}^3$ than in ${\mathbb R}^2$.

\bigskip
\noindent {\bf ACKNOWLEDGEMENTS:}
\medskip

S.T. thanks the Institute for Advanced Study for its hospitality during his stay there. This work was supported by the Division of Mathematical Sciences at the National Science Foundation under Award Number DMS-0804431 and by the MRSEC Program of the National Science Foundation under Award Number DMR-0820341.

\appendix
\section{Description of the algorithm}
\label{algorithm}
The algorithm employed to find the DLP optimal packings presented operates in two steps repeated iteratively in succession. In the first step, we use a method in nonlinear programming often called an augmented Lagrangian method \cite{Ruszczynski2006} to find a local minimum of a specially formulated problem in $Nd + 1$ variables, where the first $Nd$ variables correspond to coordinates of the centers of $N$ spheres in $d$ dimensions. In the second step, the configuration of $N$ spheres found in the first step is spatially repositioned using a random number generator. Keeping track of the least local minimum, the two steps are repeated iteratively until (presumably) no further improvement on the least minimum can be achieved \cite{endnote2}.

The nonlinear programming algorithm used in the first step seeks to solve a problem posed for $Nd + 1$ variables with $2N\! +\, _N\!C_2$ inequality constraints, where $_N\!C_2$ is shorthand for the standard combinatorial formula. Calling the first $Nd$ variables $x_{ik}$ with $i=1\dots N$ and $k=1\dots d$ and the last variable $\omega$, the problem may be posed as follows:
\begin{eqnarray}
\min \omega \,\,\,\, s.t. \notag \\
1-\sum_{k=1}^dx_{ik}^2 \leq 0 \qquad & \forall & \, i:\,\,1\dots N \notag \\
-\omega + \sum_{k=1}^dx_{ik}^2 \leq 0 \qquad & \forall & \, i:\,\,1\dots N \notag \\
1-\sum_{k=1}^d(x_{ik}-x_{jk})^2 \leq 0 \qquad & \forall & \, i,j:\,\,1\dots N,\,\, i<j. \label{CSPnonlinearForm}
\end{eqnarray}

The constraints in (\ref{CSPnonlinearForm}) are split into three categories; the first category, containing $N$ constraints, requires that none of the centers of the spheres move within distance unity of the origin, i.e., that none of the $N$ spheres overlap the fixed central sphere. The second category, also containing $N$ constraints, requires that the difference between the squared distance from the origin to the centers of each of the $N$ spheres and the objective function, simply the independent variable $\omega$, be less than or equal to zero. Since $\omega$ is minimized, this condition sets $\omega$ equal to $R^2$, i.e., the greatest of the squared distances from the origin to the any of the $N$ sphere centers. The third category, containing $_N\!C_2$ constraints, requires that none of the $N$ spheres overlap.

An augmented Lagrangian (AL) method is employed to attempt solutions to the problem \ref{CSPnonlinearForm}. In brief, an AL method is an iterative process designed to minimize a function that is the Lagrangian of (\ref{CSPnonlinearForm}) augmented with a quadratic penalty function. The augmented Lagrangian is written,
\begin{equation}
L_{\gamma}({\bf x},{\bf \lambda}) = \omega + \frac{\gamma}{2}\sum_{l=1}^m\left[max\left(0,c_l({\bf x}) + \frac{\lambda_l}{\gamma}\right)\right]^2 - \frac{1}{2\gamma}\sum_{l=1}^m\lambda_l^2,
\label{augLagr}
\end{equation}
where the $\{\lambda_l\}$ are the Lagrange multipliers, $\gamma$ is the penalty parameter (often denoted by $\rho$), and the $c_l({\bf x})$ (often denoted by $g_l({\bf x})$)  are the $m = 2N\! +\, _N\!C_2$ inequality constraints. The function (\ref{augLagr}) is minimized iteratively over ${\bf x}$, yielding $\hat{{\bf x}}_p$ a local minimum of $L_{\gamma}({\bf x}, {\bf \lambda}_p)$, with ${\bf \lambda}_p$ fixed. For each successive $\hat{{\bf x}}_p$ that is found, a new estimate for ${\bf \lambda}_p$ is made based on the violated constraints, i.e., the constraints in (\ref{CSPnonlinearForm}) with positive values. Concurrently, the penalty parameter $\gamma$ is increased by a pre-specified multiple if the cumulative squared violation, or total error, is not a set amount smaller than the previous total error. Eventually, $\gamma$ reaches a value such that the total error is smaller than a specified (small) tolerance, at which point the algorithm terminates.

In the version of the AL method we use to find putative DLP optimal packings, the function $L_{\gamma}({\bf x},{\bf \lambda})$ in each iteration is minimized using a conjugate gradient method with a directional minimizing algorithm that employs cubic interpolation. Though the conjugate gradient method can only guarantee a global minimum when the function to be minimized is quadratic, as $L_{\gamma}({\bf x}, {\bf \lambda})$ is quartic in ${\bf x}$, the method is nonetheless efficient. It is important to note also that a directional minimizing algorithm employing cubic interpolation does not produce step sizes that would guarantee a global minimum through the conjugate gradient method even for a quadratic function; this feature is key in producing an AL algorithm that does not easily become trapped in local minima.

Essentially, the AL method employed seeks to minimize $\omega = R^2$ iteratively with an increasing penalty for overlap between spheres. This can be seen as beginning with $N$ permeable spheres, squeezing them together within a spherical boundary of radius $R$, and then iteratively decreasing their permeability such that they force the boundary outward. The AL method also does not guarantee a global minimum of (\ref{CSPnonlinearForm}), though for large enough values of $\gamma$ in certain problems, $\gamma > \gamma_0$, a local saddle point of (\ref{augLagr}) is guaranteed to exist \cite{Ruszczynski2006}. For more detailed information on the augmented Lagrangian and conjugate gradient methods, we refer the reader to one of the many texts on the subject of nonlinear programming, such as \cite{Ruszczynski2006}.

The local minimum of (\ref{CSPnonlinearForm}) found by the AL method for large enough $\gamma$ is dependent not only on the algorithm parameters but also strongly on the initial conditions for ${\bf x}$ and ${\bf \lambda}$. Accordingly, the procedure we use to find global minima begins with a variety of initial conditions for ${\bf x}$, from disks positioned randomly according to the Poisson distribution inside or on the surface of a disk of a larger radius to disks positioned with centers on the sites of the triangular lattice. After each iteration of the AL method, a subset of the disk centers arranged as the local minimum in ${\bf x}$ are moved radially inward by a random amount not exceeding their initial distance from the origin. Immediately after, these same disks are rotated through a random angle such that they are no farther than distance unity from their previous position. This `shuffled' configuration is used as the initial conditions for the first $Nd$ variables of ${\bf x}$ in the next AL iteration, while the final variable of ${\bf x}$ is set to the greatest squared distance from the origin to any of the $N$ disk centers. The initial ${\bf \lambda}$ at every iteration is set to zero to help keep the AL algorithm from becoming stuck in a local minimum. 

In our trials, generally no more than $50$ iterative AL and shuffling steps were necessary to find a DLP optimal packing, even for large numbers of disks (for small numbers of disks, as few as $1$ or $10$ steps was often sufficient). To support the conjecture that the minima found are indeed global minima, we repeated each procedure of $50$ iterations as many as $20$ times, changing the governing parameters of the AL algorithm and employing the different initial conditions discussed above. For the vast majority of $N$ in the packings presented in this work, the same least local minima were found in all or the majority of the $20$ repetitions performed.

\section{Characteristics of the DLP optimal packings presented}
\label{tables}
The following table lists the putative $R_{min}(N)$, calculated with accuracy to at least $10^{-6}$ diameter units, for each packing presented. For comparison with the triangular lattice, alongside each $R_{min}(N)$ is listed the smallest radius triangular lattice shell enclosing the centers of at least $N$ disks (not including the central disk). Point group symmetries are additionally listed, as determined also with accuracy to at least $10^{-6}$ diameter units.

\begin{center}
\tablefirsthead{
\multicolumn{5}{c}{{\bf DLP optimal packing characteristics}} \\
\hline
\hline
Number of disks \,\, & Figure \,\, & \,\, $R_{min}(N)$ \,\, & \,\, Tri. latt. shell \,\, & \,\, Point group \,\,  \\
\hline
}
\tablehead{
\hline
\multicolumn{5}{l}{\textit{continued from previous page}} \\
\hline
Number of disks \,\, & Figure \,\, & \,\, $R_{min}(N)$ \,\, & \,\, Tri. latt. shell \,\, & \,\, Point group \,\,  \\
\hline}
\tabletail{
\hline
\multicolumn{5}{r}{\textit{continued on next page}} \\
\hline}
\tablelasttail{
\hline}
\begin{supertabular}{c l c c c}
10 & \ref{10opt} (top) & 1.618034 & 1.732051 & $C_{2v}$ \\
10 & \ref{10opt} (center) & 1.618034 & 1.732051 & $C_{2v}$ \\
10 & \ref{10opt} (bottom) & 1.618034 & 1.732051 & $D_{5h}$ \\
11 & \ref{oddSyms} (center) & 1.685854 & 1.732051 & $C_{2v}$ \\
15 & \ref{DLP15pic} & 1.873123 & 2 & $D_{5h}$ \\
24 & \ref{circCavities} (top) & 2.425256 & 2.645751 & $D_{3h}$ \\
25 & \ref{oddSyms} (top) & 2.497212 & 2.645751 & $D_{5h}$ \\
32 & \ref{oddSyms} (bottom) & 2.794164 & 3 & $D_{2h}$ \\
40 & \ref{eyePackings} (top) & 3.136712 & 3.464102 & $D_{2h}$ \\
45 & \ref{circCavities} (center) & 3.374023 & 3.605551 & $C_1$ \\
46 & \ref{noSym} (bottom) & 3.414304 & 3.605551 & $C_1$ \\
59 & \ref{noSym} (top) & 3.824374 & 4 & $C_1$ \\
60 & \ref{curvedHex} (top) & 3.830649 & 4 & $C_{6h}$ \\
66 & \ref{eyePackings} (bottom) & 4.104997 & 4.358899 & $C_1$ \\
80 & \ref{noSym} (center) & 4.514170 & 4.582576 & $C_1$ \\
84 & \ref{wedgeHexA} (top) & 4.581556 & 4.582576 & $C_i$ \\
90 & \ref{curvedHex} (center) & 4.783386 & 5 & $C_{6h}$ \\
95 & \ref{circCavities} (bottom) & 4.958096 & 5.196152 & $C_1$ \\
120 & \ref{wedgeHexA} (center) & 5.562401 & 5.567764 & $C_i$ \\
126 & \ref{curvedHex} (bottom) & 5.736857 & 6 & $C_{6h}$ \\
162 & \ref{wedgeHexA} (bottom) & 6.539939 & 6.557439 & $C_i$ \\
168 & \ref{bigDiff} (top) & 6.680013 & 6.928203 & $C_1$ \\
198 & \ref{wedgeHexB} (top) & 7.201130 & 7.211103 & $C_i$ \\
264 & \ref{bigDiff} (center) & 8.417769 & 8.544004 & $C_1$ \\
270 & \ref{bigDiff} (bottom) & 8.497744 & 8.660254 & $C_1$ \\
312 & \ref{wedgeHexB} (bottom) & 9.141107 & 9.165151 & $C_i$ \\
348 & \ref{manySpheres} & 9.620709 & 9.643651 & $D_{6h}$ \\
\hline
\end{supertabular}
\end{center}


\begin{thebibliography}{32}
\bibitem[{end({\natexlab{a}})}]{endnote3}
\bibinfo{note}{For all putative optimal packings presented in this paper for
  $N$ up to $N=168$, accuracy is to at least $10^{-8}$ diameters, while for
  some $N>168$, an accuracy to at least $10^{-6}$ diameters was used to reduce
  computation time. These accuracy values apply to all $R_{min}(N)$, the
  contact distance between disks in contact and the tolerance used in
  identifying symmetry elements in order to assign point groups. Additionally,
  we have used the convention of placing dots after a $R_{min}(N)$ value, such
  as with $R_{min}(15) = 1.873123\dots$, to imply that the complete
  $R_{min}(N)$ value may be calculated analytically using trigonmetry (though
  we do not present these calculations here). This is the case for every
  packing presented with point group other than $C_1$, as exploitation of
  existing symmetries in symmetric optimal packings allows explicit calculation
  of $R_{min}(N)$.}
\bibitem[{end({\natexlab{c}})}]{endnote5}
\bibinfo{note}{In practice, for $d=2$ we find that the only $N > 6$ for which
  $R_{min}(N) = R_{min}(N-1)$ is $N=18$.}
\bibitem[{end({\natexlab{d}})}]{endnote6}
\bibinfo{note}{Note that this definition allows for a number of spheres to have
  only a fraction of their volume included in the window.}
\bibitem[{end({\natexlab{e}})}]{endnote2}
\bibinfo{note}{Although for this work we sought DLP optimal configurations only
  for packings of identical nonoverlapping spheres, the two step process is
  versatile and can be adapted to attempt solutions to any packing problem
  seeking the maximum packing fraction of a finite number of identical or
  differentiated nonoverlapping objects.}
\end{thebibliography}
\end{document}